\documentclass[reprint,aps,prb,amsmath,amssymb,amsfonts,dvipdfmx]{revtex4-1}
\usepackage{graphicx}
\usepackage{color}
\usepackage{braket,bm}
\usepackage{dcolumn}
\begin{document}

\title{Randomness-induced quantum spin liquid behavior in the $s=\frac{1}{2}$ random $J_1$-$J_2$ Heisenberg antiferromagnet on the square lattice}
\author{Kazuki Uematsu}
\email[]{uematsu@spin.ess.sci.osaka-u.ac.jp}
\affiliation{Department of Earth and Space Science, Graduate School of Science, Osaka University, Toyonaka, Osaka 560-0043, Japan}
\author{Hikaru Kawamura}
\email[]{kawamura@ess.sci.osaka-u.ac.jp}
\affiliation{Department of Earth and Space Science, Graduate School of Science, Osaka University, Toyonaka, Osaka 560-0043, Japan}


\begin{abstract}
We investigate the ground-state and the finite-temperature properties of the bond-random $s=1/2$ Heisenberg model on a square lattice with frustrating nearest- and next-nearest-neighbor antiferromagnetic interactions, $J_1$ and $J_2$, by the exact diagonalization and the Hams--de Raedt methods. The ground-state phase diagram of the model is constructed in the randomness versus the frustration ($J_2/J_1$) plane, with the aim of clarifying the effects of randomness and frustration in stabilizing a variety of phases. We find that the randomness induces the gapless quantum spin liquid (QSL)-like state, the random-singlet state, together with the spin-glass state in a certain range of parameter space. The spin-glass state might be stabilized by employing the lattice directional degrees of freedom associated with the stripe-type magnetic order of the regular model. Possible implications to recent experiments on the square-lattice mixed-crystal antiferromagnet Sr$_2$Cu(Te$_{1-x}$W$_{x}$)O$_6$ exhibiting the gapless QSL-like behaviors are discussed.
\end{abstract}

\pacs{}

\maketitle

\section{Introduction}
\label{sec:intro}

The quantum spin liquid (QSL) state is an exotic state of magnets, not accompanying any magnetic long-range order (LRO) nor exhibiting any spontaneous symmetry breaking down to low temperatures. The related issue has extensively been studied since the proposal of the resonating valence bond state by Anderson in the 1970s.\cite{AndersonRVB} In the early 2000s, the QSL-like behaviors were experimentally observed in some frustrated quantum antiferromagnets. For example, the organic triangular-lattice salts $\kappa$-(ET)$_2$Cu$_2$(CN)$_3$ \cite{ET, ET-Kanoda,ET-Matsuda,ET-Jawad,ET-Sasaki} and EtMe$_3$Sb[Pd(dmit)$_2$]$_2$ \cite{dmit,dmit-Matsuda,dmit-Kato,dmit-Jawad} were found not to exhibit any magnetic order down to very low temperature, with gapless behaviors characterized by the low-temperature specific heat linear in the absolute temperature $T$.\cite{ET-Kanoda,ET-Matsuda,dmit-Matsuda,dmit-Kato} The inorganic kagome-lattice compound herbertsmithite ZnCu$_3$(OH)$_6$Cl$_2$ is also a well-studied example of the QSL.\cite{Shores,Helton,Olariu,Freedman,Han,Imai}  This kagome material was reported to exhibit gapless QSL behaviors with broad features in the dynamical spin structure factor,\cite{Helton,Olariu,Han} whereas a recent NMR study reported a nonzero spin gap.\cite{Imai} There are so many theoretical suggestions related to the observed QSL-like behaviors, while the true nature of the experimentally observed QSL behaviors is still not fully understood and is under debate even now.

Recently, the randomness was invoked as a key ingredient in inducing the QSL-like behaviors in many of experimentally observed QSL magnets.\cite{Singh,Watanabe,Kawamura,Shimokawa,Savary,Uematsu,Kimchi,Wu,Kimchi2,Lin-Sandvik} One of the present authors (H.K.) and collaborators have claimed, on the basis of a series of numerical computation on the geometrically frustrated bond-random quantum Heisenberg models, that the QSL-like behaviors observed in triangular-lattice organic salts and kagome-lattice herbertsmithite might be the randomness-induced one, ``the random-singlet state.'' \cite{Watanabe,Kawamura,Shimokawa} A similar nonmagnetic state was also proposed for the kagome herbertsmithite by Singh based on the site-random kagome Heisenberg model. \cite{Singh}

 The random-singlet state proposed in Refs. \onlinecite{Watanabe,Kawamura,Shimokawa} for the 2D frustrated quantum magnets is a gapless QSL-like state where spin-singlets of varying strength are formed randomly distributed in space, adjusting to randomly modulated exchange interactions ${J_{ij}}$ and locally resonating between energetically degenerate singlet coverings.

 The origin of the randomness might be either (i) an extrinsic or conventional quenched randomness such as impurities, defects and intersite disorder in herbertsmithite or (ii) an intrinsic ``effective'' randomness, i.e., an inhomogeneity dynamically self-generated in the spin sector via the coupling to other degrees of freedom in magnets such as charge or dielectric degrees of freedom as in case of organic salts. The random-singlet state of Refs. \onlinecite{Watanabe,Kawamura,Shimokawa} is characterized by the $T$-linear specific heat, the gapless susceptibility with an intrinsic Curie tail, and broad features of the dynamical spin structure factor, which are well consistent with the experimental features of $\kappa$-(ET)$_2$Cu$_2$(CN)$_3$, EtMe$_3$Sb[Pd(dmit)$_2$]$_2$, and ZnCu$_3$(OH)$_6$Cl$_2$.

 The random-singlet-like state, vaguely defined here as a randomness-induced singlet-based nonmagnetic state, was discussed for some time in the literature, mainly in the context of random 1D magnets, \cite{DusguptaMa,Fisher} but also of loosely coupled  spins in dilute magnetic semiconductors. \cite{BhattLee} How the random-singlet state discussed in Refs. \onlinecite{Watanabe,Kawamura,Shimokawa} in the context of frustrated 2D quantum magnets resembles or differs from the ones discussed in 1D and in dilute magnetic semiconductors is not entirely clear at the present stage, and needs further clarification.

 In case of frustrated quantum magnets in 2D, it was recently shown by the present authors that the random-singlet state is generically found not only in geometrically frustrated lattices such as triangular and kagome lattices but also for geometrically unfrustrated lattices such as a honeycomb lattice, once the frustration is introduced via, e.g., the competition between the nearest-neighbor and the next-nearest-neighbor interactions $J_1$ and $J_2$.\cite{Uematsu} This might imply that the gapless QSL-like behaviors could be realized in a wide class of quantum 2D magnets possessing a certain amount of randomness without fine-tuning the interaction parameters.

\begin{figure}
  \begin{center}
    \includegraphics[clip,width=0.5\hsize]{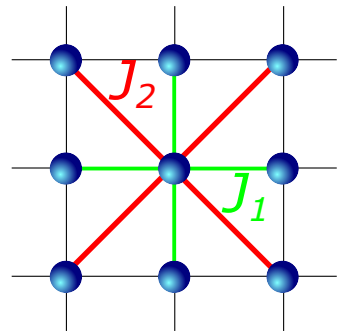}\par
    \caption{(Color online) Illustration of the square lattice, with the nearest-neighbor interaction $J_1$ (green), and the next-nearest-neighbor interaction $J_2$ (red).}
    \label{fig:square}
  \end{center}
\end{figure}

 Under such circumstances, to further clarify the nature of the randomness-induced QSL-like state, especially how the state is generic in randomly frustrated 2D quantum magnets, we study in the present paper the random $s=1/2$ Heisenberg model on the square lattice with the competing nearest- and next-nearest-neighbor antiferromagnetic interactions $J_1$ and $J_2$ (see Fig. \ref{fig:square}). In the model, one can freely adjust the extents of randomness and frustration by tuning the parameters $\Delta$ (to be defined below in Sec. \ref{sec:modelmethod}) and $J_2/J_1$, as was done in Ref. \onlinecite{Uematsu} for the honeycomb-lattice model. In the square lattice, the number of nearest-neighbor bonds is four, more than that on the honeycomb lattice, three, so that the magnetically ordered states is expected to be more robust against fluctuations compared to the honeycomb-lattice case.

\begin{figure}
  \begin{center}
    \begin{tabular}{c}
      \begin{minipage}{0.5\hsize}
        \begin{center}
          \includegraphics[clip,width=0.8\hsize]{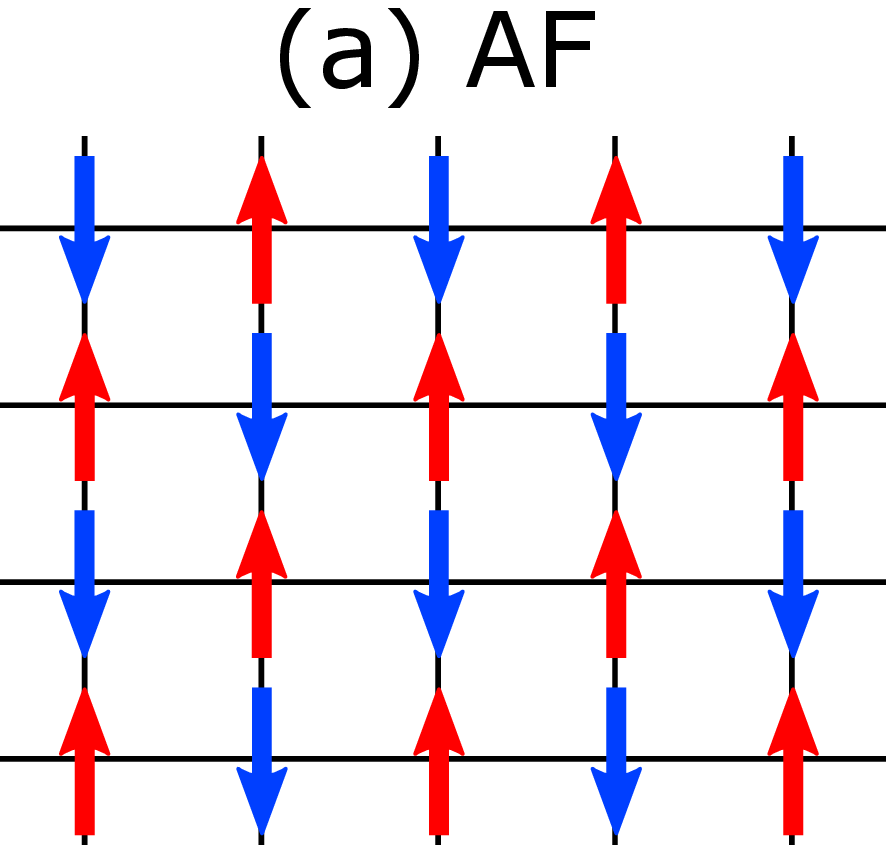}\par
        \end{center}
      \end{minipage}
      \begin{minipage}{0.5\hsize}
        \begin{center}
          \includegraphics[clip,width=0.8\hsize]{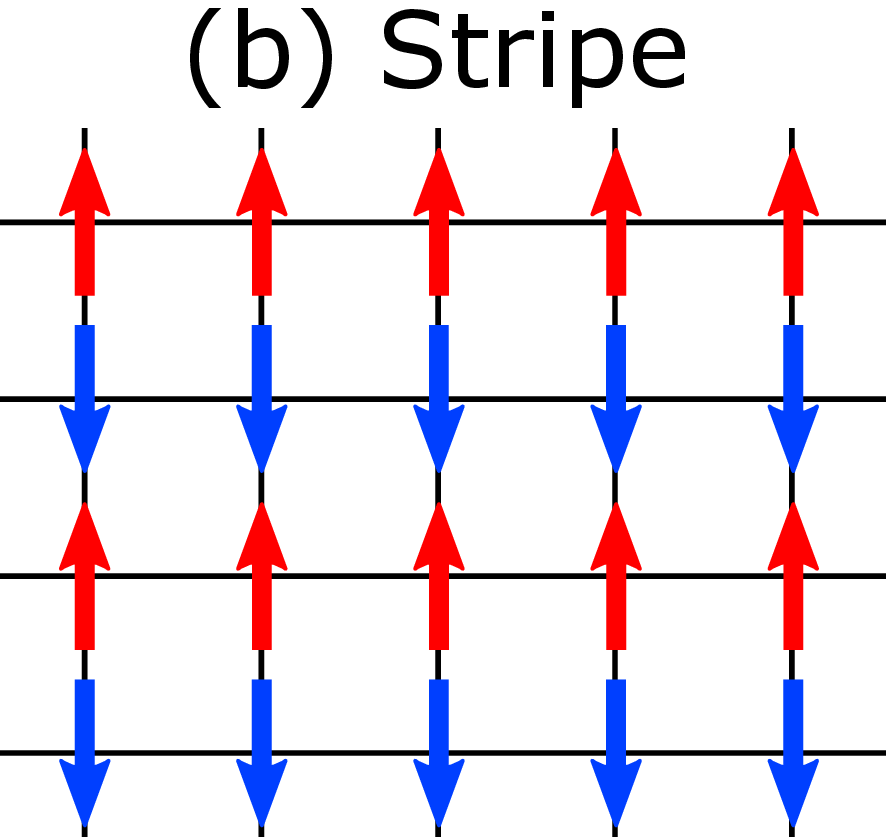}\par
        \end{center}
      \end{minipage}
    \end{tabular}
    \caption{(Color online) Magnetically ordered states of the $J_1$-$J_2$  Heisenberg antiferromagnet on the square lattice; (a) the two-sublattice AF state, and (b) the stripe-ordered state.}
    \label{fig:magnetic}
  \end{center}
\end{figure}

 The phase structure of the corresponding {\it regular} model, i.e., the $s=1/2$ $J_1$-$J_2$ Heisenberg model on the square lattice, has long been studied by various numerical methods, \cite{Schmidt-Thalmeier} including the exact diagonalization (ED) method,\cite{Shulz,Mambrini,Richter} the density matrix renormalization group method,\cite{Balents,Sheng-DMRG,Wang-DMRG} the tensor network state (TNS) algorithm,\cite{Verstraete-PEPS,Yu,Wang-Verstraete-PEPS,Wang-Verstraete-TPS,Sheng-PEPS} the variational Monte-Carlo method,\cite{Imada} and the cluster mean-field theory.\cite{Ren} For $J_2/J_1\lesssim0.4$, the ground state is the standard two-sublattice antiferromagnetic (AF) state as illustrated in Fig. \ref{fig:magnetic} (a), while it is the stripe-ordered state as illustrated in Fig. \ref{fig:magnetic} (b) for $J_2\gtrsim0.6$. For intermediate values of $0.4 \lesssim J_2/J_1\lesssim 0.6$, some kind of nonmagnetic state is likely to arise. The nature of this nonmagnetic state and the precise location of the borderline values of $J_2/J_1$ between the phases still remains controversial. Most of the theoretical studies\cite{Mambrini,Richter,Balents,Verstraete-PEPS,Sheng-PEPS,Yu} suggested the nonmagnetic state to be a gapped one such as the columnar valence bond crystal (VBC),\cite{Verstraete-PEPS,Sheng-PEPS} the plaquette VBC,\cite{Mambrini,Yu} and the $Z_2$ spin liquid,\cite{Balents} whereas one of the TNS studies suggested it to be a gapless one.\cite{Wang-Verstraete-PEPS} Some other studies suggested that the nonmagnetic state is divided into the gapless state at $J_2\lesssim0.5$ and the gapped VBC state at $J_2\gtrsim0.5$.\cite{Wang-DMRG,Sheng-DMRG,Imada} 

 For the square-lattice magnets, experimental reports of the QSL-like behavior have been scarce. One interesting example was recently reported, however, in the mixed-crystal antiferromagnet Sr$_2$Cu(Te$_{1-x}$W$_{x}$)O$_6$.\cite{Mustonen,Watanabe-Tanaka,Mustonen2} This mixed-crystal magnet for $x=0.5$ showed no indication of spin freezing down to 19mK, accompanied with a significant $T$-linear term in the specific heat. Indeed, its isostructural end materials Sr$_2$CuTeO$_6$ and Sr$_2$CuWO$_6$ exhibit the standard two-sublattice AF (Neel AF) order and the stripe (columnar AF) order, respectively. Since Sr$_2$CuTeO$_6$ has a predominant $J_1$ interaction ($J_2/J_1\sim0$) \cite{Watanabe-Tanaka,Babkevich}, and Sr$_2$CuWO$_6$ a predominant $J_2$ interaction ($J_2/J_1\sim4$-$8$)\cite{Watanabe-Tanaka,Walker}, one may expect that the $J_2/J_1\sim0.5$ region where the QSL is stabilized in the regular $J_1$-$J_2$ model might be realized in Sr$_2$Cu(Te$_{1-x}$W$_{x}$)O$_6$. Meanwhile, a significant amount of quenched randomness is expected in Sr$_2$Cu(Te$_{1-x}$W$_{x}$)O$_6$ due to the obvious reason of the mixed nature of Te and W. Another possible scenario then might be that the experimentally observed QSL state is the random-singlet state. This expectation provides another motivation for our present study.

 The organization of this paper is as follows. In Sec. \ref{sec:modelmethod}, we introduce our model, i.e., the bond-random $s=1/2$ $J_1$-$J_2$ Heisenberg model on the square lattice, and explain the computational methods employed. The ground-state properties of the model are studied by means of the ED method in Sec. \ref{sec:random}. The ground-state phase diagram is constructed in the frustration ($J_2/J_1$) versus the randomness plane, and the properties of each phase are clarified. The finite-temperature properties of the model are studied by means of the Hams-de Raedt method in Sec. \ref{sec:finitemp}. Section \ref{sec:summary} is devoted to summary and discussion. In the Appendix, we treat the $s=1/2$ $J_1$-$J_2$-$J_3$ Heisenberg model (with $J_2=J_3$) to get further information about the relative stability of the random-singlet and the spin-glass states.

\begin{figure*}[t]
  \begin{center}
    \includegraphics[clip,width=\hsize]{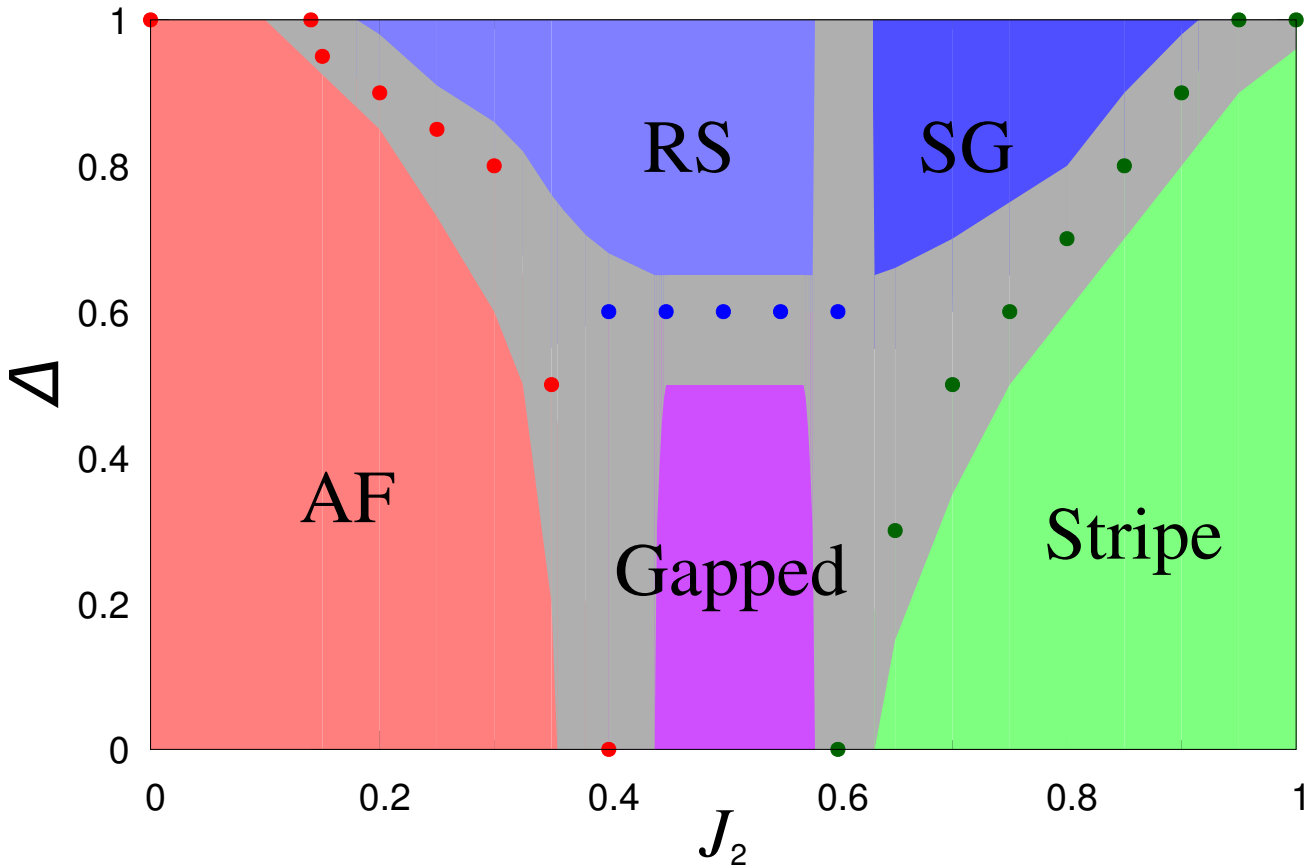}
    \caption{(Color online) Ground-state phase diagram of the $s=1/2$ bond-random $J_1$-$J_2$ ($J_1$ and $J_2$ are both antiferromagnetic) Heisenberg model on the square lattice in the frustration ($J_2$) versus the randomness  ($\Delta$) plane. ``AF'', ``Stripe'', and ``Gapped'' represent the two-sublattice antiferromagnetic state, the stripe-ordered state, and the nonmagnetic state with a finite spin gap, while ``RS'' and ``SG'' represent the random-singlet state and the spin-glass state, respectively. The red and green points denote the transition points estimated from the AF and the stripe order parameters, while the blue points denote those estimated from the spin gap.} 
    \label{fig:phase-random}
  \end{center}
\end{figure*}

\section{The model and the method}
\label{sec:modelmethod}

We consider the bond-random $s=1/2$ isotropic Heisenberg model on the square lattice with the AF nearest-neighbor and next-nearest-neighbor interactions $J_1>0$ and $J_2>0$. The Hamiltonian is given by
\begin{align}
\mathcal{H}=J_1\sum_{\Braket{i,j}}j_{ij}{\bm S}_i\cdot{\bm S}_j +
J_2\sum_{\Braket{\Braket{i,j}}} j_{ij}{\bm S}_i\cdot{\bm S}_j,
\label{eq:hamiltonian}
\end{align}
where ${\bm S}_i=(S_i^x, S_i^y, S_i^z)$ is the $s=1/2$ spin operator at the $i$-th site on the square lattice, the sums $\Braket{i,j}$ and $\Braket{\Braket{i,j}}$ are taken over all nearest-neighbor and next-nearest-neighbor pairs on the lattice, while $j_{ij}\ge 0$ is the random variable obeying the bond-independent uniform distribution between $[1-\Delta, 1+\Delta]$ with $0\leq \Delta\leq 1$.  Periodic boundary conditions are applied. Hereafter, we put $J_1=1$ and $J_2/J_1=J_2>0$. Then, the parameter $J_2$ represents the degree of frustration borne by the competition between $J_1$ and $J_2$. Our present choice of the bond-independent uniform distribution for $j_{ij}$ is just for simplicity, whereas, in real materials, the distribution could be more complex and correlated. The parameter $\Delta$ represents the extent of the randomness: $\Delta=0$ corresponds to the regular case and $\Delta=1$ to the maximally random case. The extent of the randomness $\Delta$ is taken to be common between $J_1$ and $J_2$, again just for simplicity. Note that, by tuning the parameters $\Delta$ and $J_2$, we can control the degrees of both the randomness and the frustration independently as was done in the honeycomb-lattice model.\cite{Uematsu}

 The ground-state properties of the model are computed by the ED Lanczos method. We treat finite-size clusters with the total number of spins $N$ up to $N\leq32$ (all even-$N$ samples with $8\leq N\leq 32$). All clusters studied are commensurate with the two-sublattice AF order illustrated in Fig. \ref{fig:magnetic} (a). The clusters of $N=8$, 12, 16, 20, 24, 28, and 32 are commensurate with the stripe order of Fig. \ref{fig:magnetic} (b), among which $N=8$, 16, 20, and 32 possess the fourfold rotational symmetry of the bulk square lattice. More generally, the clusters of $N=8$, 10, 16, 18, 20, 26, and 32 possess the fourfold rotational symmetry of the bulk square lattice.

 The number of independent bond realizations used in the configurational or sample average is $N_s=100$, 50, 25, 16, and 10 for $N=8$--$24$, 26, 28, 30, and 32 for the order parameter, the spin gap and the static spin structure factor,  whereas $N_s=100$, 100, and 25 for $N=16,20$, and 32 for the dynamical spin structure factor, respectively. Error bars are estimated from sample-to-sample fluctuations.

\begin{figure}
  \begin{center}
    \includegraphics[clip,width=\hsize]{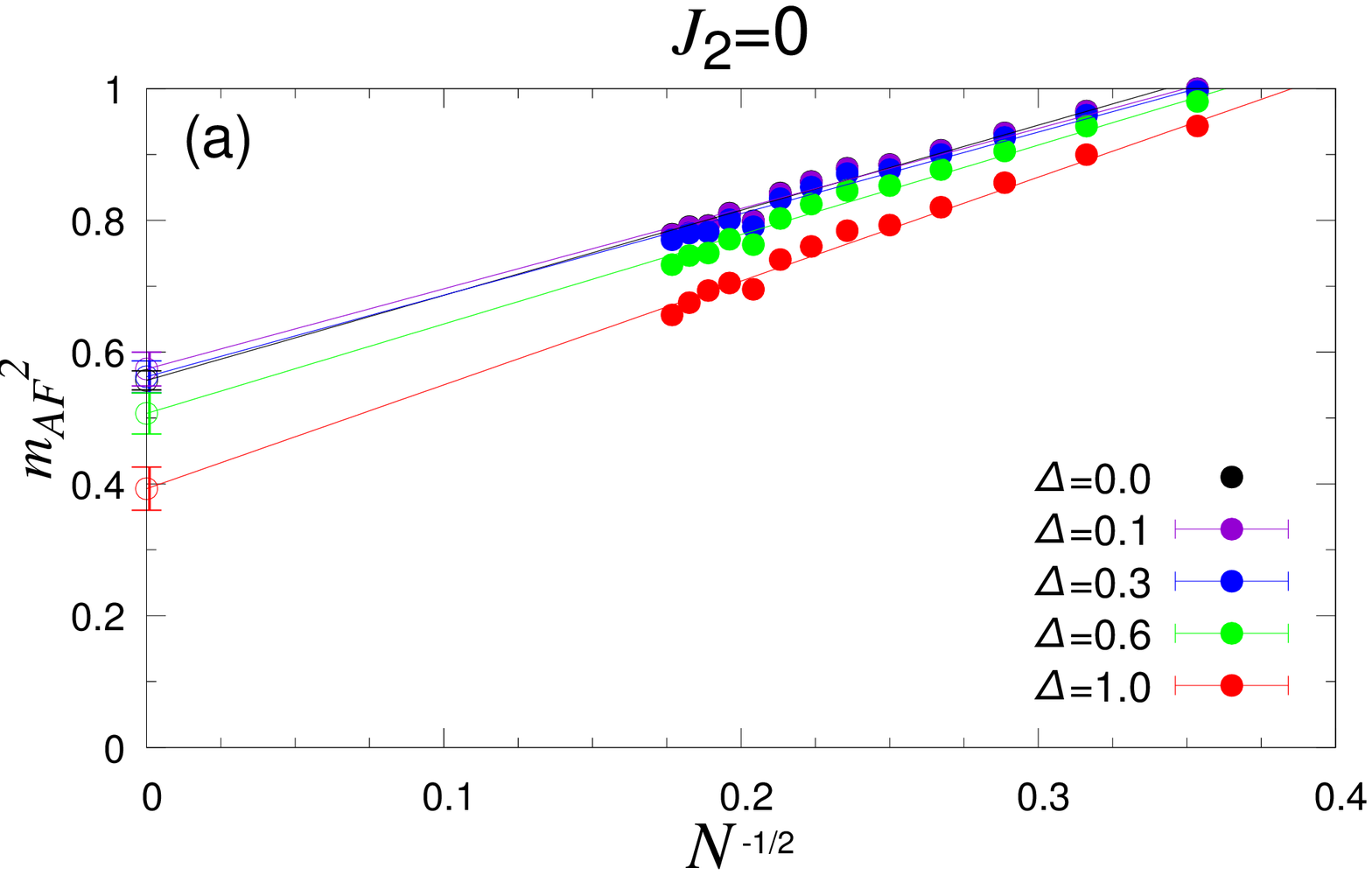}
    \includegraphics[clip,width=\hsize]{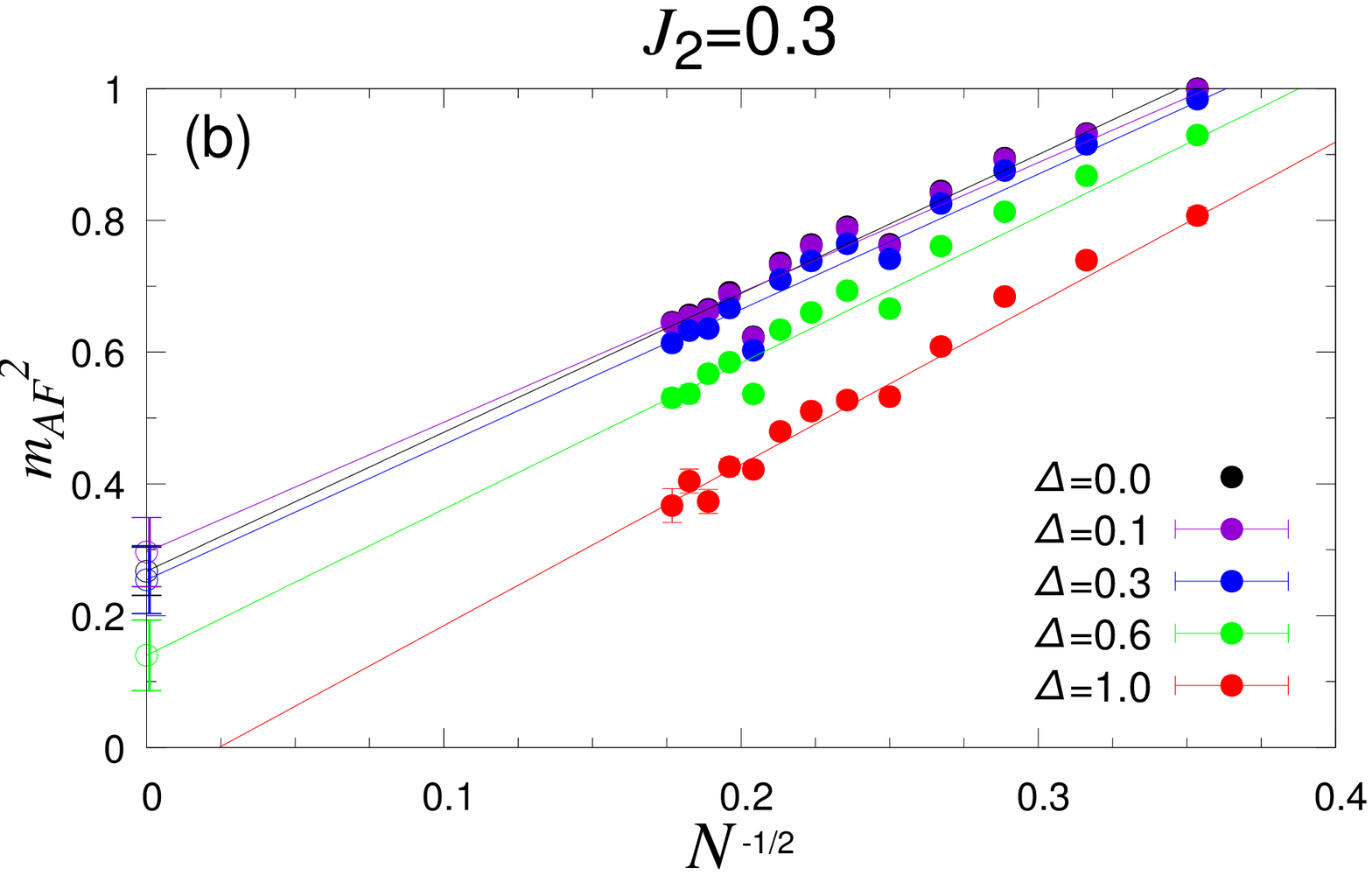}
    \caption{(Color online) The squared two-sublattice AF order parameter $m_{AF}^2$ plotted versus $1/\sqrt{N}$ for various values of $\Delta$, for (a) $J_2=0$ and (b) $J_2=0.3$. The lines are linear fits of the data.} 
    \label{fig:mAF}
  \end{center}
\end{figure}
\begin{figure}
  \begin{center}
    \includegraphics[clip,width=\hsize]{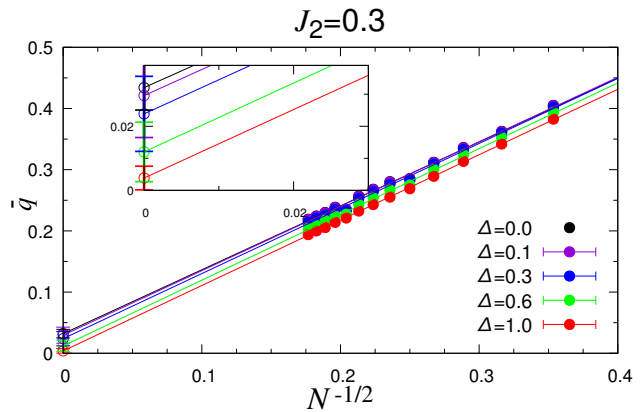}
    \caption{(Color online) The spin freezing parameter $\bar{q}$ plotted versus  $1/\sqrt{N}$ for various values of $\Delta$ for $J_2=0.3$. The lines are linear fits of the data. The inset is a magnified view of the large-$N$ region.}
    \label{fig:qb1}
  \end{center}
\end{figure}
\begin{figure}
  \begin{center}
    \includegraphics[clip,width=0.7\hsize]{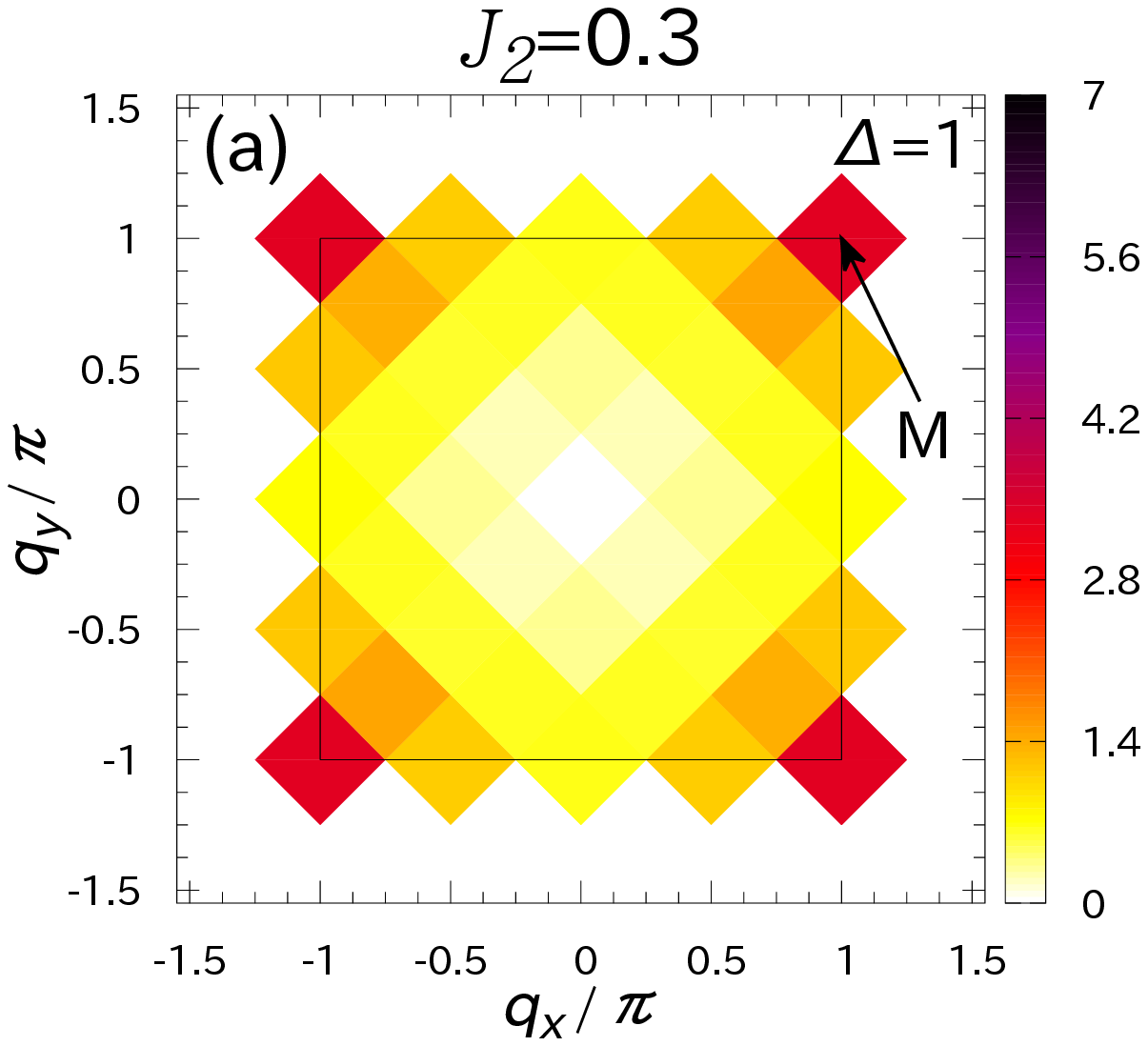}
    \includegraphics[width=\hsize]{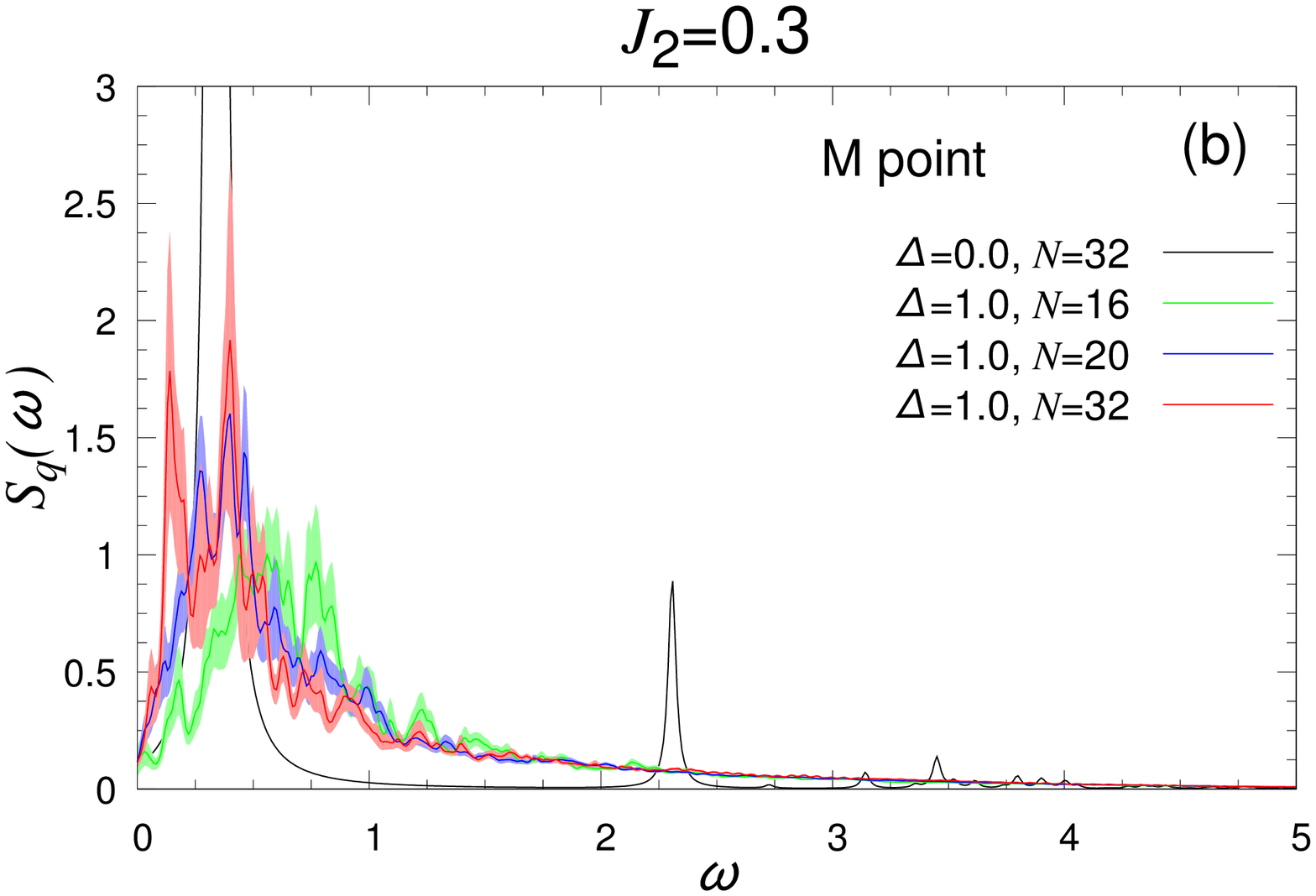}    
\caption{(Color online) (a) Intensity plots of the static spin structure factor $S_{\bm q}$ for the size $N=32$, and (b) the $\omega$-dependence of the dynamical spin structure factor $S_{\bm q}(\omega)$ computed at the M point for the sizes $N=16, 20$ and 32, for $J_2=0.3$ and $\Delta=1$. In (a), the solid line exhibits the boundary of the first Brillouin zone. In (b), the data of the regular model of $\Delta=0$ are also shown for comparison.}
\label{fig:SSF1}
  \end{center}
\end{figure}

 The finite-temperature properties are computed by the Hams--de Raedt method, \cite{HamsRaedt} where the thermal average is replaced by the average over a few ``pure states'' produced via the imaginary time-evolution of initial random vectors. The method enables us to calculate various finite-temperature properties at nearly the same computational cost as that of the Lanczos method. Our finite-temperature computation is performed for the size $N=24$, where the averaging is made over 30 initial vectors and 50 independent bond realizations. Error bars of physical quantities are estimated from the scattering over both samples and initial states by using the bootstrap method.

\section{The ground-state properties and the phase diagram}
\label{sec:random}

\subsection{Phase diagram}

In this section, we present our numerical results on the  bond-random $s=1/2$ $J_1$-$J_2$ Heisenberg model on the square lattice in the region of $0\leq J_2\leq 1$. In Fig. \ref{fig:phase-random}, we first show our main result, i.e., the ground-state phase diagram in the frustration ($J_2$) versus the randomness ($\Delta$) plane. The $\Delta=0$ line corresponds to the phase diagram of the regular model studied by the previous authors. In fact, when the randomness $\Delta$ is sufficiently weak, the phase diagram turns out to be qualitatively similar to that of the regular model.

In the phase diagram shown in Fig. \ref{fig:phase-random}, five distinct phases are identified. Three of them, i.e., the two-sublattice AF phase, the gapped nonmagnetic phase, and the stripe-ordered phase have already been identified in the regular model, while, when the strength of the randomness exceeds a critical value $\Delta_c(J_2)$, the two new phases, the random-singlet phase and the spin-glass phase, appear. The width of the parameter region of the random-singlet phase is slightly narrow compared with that of the corresponding honeycomb-lattice model.\cite{Uematsu} Instead, the spin-glass state not stabilized in the honeycomb-lattice model appears in the specific parameter region where the stripe-ordered state is stabilized in the regular limit.

The transition points are determined from the AF order parameter $m_{AF}^2$, the stripe order parameter $m_{str}^2$, the spin-gap $\Delta E$, and the spin freezing parameter $\bar{q}$. The phase boundary between the AF phase and the random-singlet phase (red points in Fig. \ref{fig:phase-random}) is determined from $m_{AF}^2$, while that between the stripe-ordered phase and the gapped phase or the random-singlet phase (green points in Fig. \ref{fig:phase-random}) is determined from $m_{str}^2$. The phase boundary between the gapped phase and the random-singlet phase (blue points in Fig. \ref{fig:phase-random}) is determined from the spin gap $\Delta E$. The spin freezing parameter $\bar{q}$ can detect any type of static spin order, even including the random one such as the spin-glass order, so that the phase boundary between the random-singlet phase and the spin-glass phase is determined from $\bar{q}$.

Below, we show our numerical data for various observables including the order parameters, the spin-gap energy, and the static and dynamical structure factors, for the $J_2$-regions of (i) $0\leq J_2\leq 0.4$, (ii) $0.4\leq J_2\leq 0.6$, and (iii) $0.6\leq J_2\leq 1$, separately.

\subsection{Region $0\leq J_2 \leq 0.4$}
First, we investigate the $0\leq J_2\leq 0.4$ region where the standard two-sublattice AF order appears in the regular limit.
As the associated AF order parameter, one can take the squared sublattice magnetization $m_{AF}^2$ defined by 
\begin{align}
  m_{AF}^2 &=\frac{1}{2}\frac{1}{\frac{N}{4}(\frac{N}{4}+1)}
  \left[\sum_{\alpha=A,B} \Braket{ \left( \sum_{i\in\alpha}{\bm S}_i
      \right)^2 } \right]_J  \nonumber \\
  &=\frac{8}{N(N+4)} \left[\sum_{\alpha}\sum_{i,j\in \alpha}
    \Braket{{\bm S}_i \cdot {\bm S}_j} \right]_J ,
\label{eq:m_AF}
\end{align}
where $\alpha=A,B$ denotes the two sublattices of the square lattice shown in Fig. 2(a), the sum over $i,j\in \alpha$ is taken over all sites $i,j$ belonging to the sublattice $\alpha$, while $\langle \cdots \rangle$ and $[\cdots]_J$ represent the ground-state expectation value (or the thermal average at finite temperatures) and the configurational average over $J_{ij}$ realizations, respectively. When the system retains a relevant LRO, $m_{\infty}^2>0$, the spin-wave analysis yields the size dependence proportional to $1/\sqrt{N}$, i.e.,
\begin{align}
m^2 = m_{\infty}^2 + \frac{c_1}{\sqrt{N}}.
\label{eq:fit-sqrt}
\end{align}

In Fig. \ref{fig:mAF}, we show the computed squared AF order parameter $m_{AF}^2$ plotted versus $1/\sqrt{N}$ for various values of randomness $\Delta$, for (a) $J_2=0$ and (b) $J_2=0.3$. For $J_2=0$, i.e., for the unfrustrated nearest-neighbor model, Fig. \ref{fig:mAF} (a) indicates that $m_\infty^2$ is always extrapolated to a nonzero positive value for any $\Delta$, demonstrating that the AF LRO is stabilized up to the maximal randomness as was already reported.\cite{Shimokawa,Laflorencie} For $J_2=0.3$, on the other hand, there exists a finite critical randomness $\Delta_c\simeq 0.8$ beyond which the AF LRO vanishes, as can be seen from Fig. \ref{fig:mAF} (b). This observation demonstrates that a certain amount of frustration is necessary to destabilize the AF LRO by introducing the randomness. \cite{Uematsu}

 To investigate the possible appearance of other types of  magnetic order, we also compute the spin freezing parameter $\bar q$
defined by
\begin{align}
\bar{q}= \frac{1}{N}
\sqrt{\left[\sum_{i,j}\Braket{{\bm S}_i\cdot{\bm S}_j}^2\right]_J}.
\label{eq:q_bar}
\end{align}
The $1/\sqrt{N}$-dependence of the computed $\bar q$ is shown in Fig. \ref{fig:qb1} for the cases of $J_2=0.3$. The inset exhibits a magnification of the larger $N$ region.
 The interest here is whether $\bar q$ could be nonzero in the parameter region without the AF LRO.\cite{Watanabe,Uematsu} As can be seen from Fig. \ref{fig:qb1},  whether the extrapolated $\bar q$ is positive or negative (zero) well correlates with the behavior of $m_{AF}^2$ shown in Fig. \ref{fig:mAF} (b), indicating that no magnetically ordered state other than the standard AF order appears in the parameter range studied. This means that the state observed at $J_2=0.3$ for a stronger randomness of $\Delta \gtrsim 0.8$ is a nonmagnetic state without any static spin order including the spin-glass order.

 To probe the wider reciprocal space, we compute the ground-state spin structure factor $S_{\bm q}$ defined by 
\begin{align}
S_{\bm q}&=\frac{1}{N}\left[ \Braket{ |{\bm S}_{\bm q}|^2}\right]_J
\nonumber \\
&=\frac{1}{N} \left[ \sum_{i,j} \Braket{{\bm S}_i\cdot{\bm S}_j}
  \cos{\left({\bm q}\cdot\left({\bm r}_i-{\bm r}_j\right)\right)}
  \right]_J,
\label{eq:SSF}
\end{align}
where ${\bm S}_{\bm q}=\sum_j {\bm S}_j e^{i{\bm q}\cdot{\bm r}_j}$ is the Fourier transform of the spin operator, and ${\bm r}_j$ is the position vector at the site $j$. The length unit is taken here to be the nearest-neighbor distance of the square lattice.
The computed $S_{\bm q}$ at $J_2=0.3$ is shown in Fig. \ref{fig:SSF1} (a) for the maximally random case of $\Delta=1$.
While the system is expected to be in the nonmagnetic random-singlet state here, one sees that a peak associated with the AF short-range order (SRO) appears at the M point located at $(\pi,\pi)$.

 We also compute the dynamical spin structure factor $S_{\bm q}(\omega)$ defined by
\begin{align}
S_{\bm q}(\omega)&=\int_{-\infty}^\infty
\left[ \Braket{\left(S_{\bm q}^z(t)\right)^\dag
  S_{\bm q}^z(0)} \right]_Je^{-i\omega t}dt \nonumber \\
&=-\lim_{\eta\to 0}\left[\frac{1}{\pi}{\rm Im}\Braket{
    (S_{\bm q}^z)^\dag\frac{1}{\omega + E_0 + i\eta -\mathcal{H}}
    S_{\bm q}^z}\right]_J ,
\label{eq:DSF}
\end{align}
where $E_0$ is the ground-state energy, and $\eta$ is a phenomenological damping factor taking a sufficiently small positive value. We employ the continued fraction method to compute $S_{\bm q}(\omega)$, \cite{Gagliano} putting $\eta=0.02$.
 In Fig. \ref{fig:SSF1} (b), we show for the case of $J_2=0.3$ the computed $\omega$-dependence of $S_{\bm q}(\omega)$ at the M point at the maximal randomness of $\Delta=1$. While a rather sharp peak is observed in the small-$\omega$ region, a very broad background component and a tail extending to larger values of $\omega$ are observed coexisting with the peak structure at smaller $\omega$. This broad feature is a characteristic of the rando-singlet state,\cite{Kawamura,Shimokawa,Uematsu} which also supports that the observed nonmagnetic state is indeed a random-singlet state.

\subsection{Region $0.4\leq J_2 \leq 0.6$}
\begin{figure}
  \begin{center}
    \includegraphics[clip,width=\hsize]{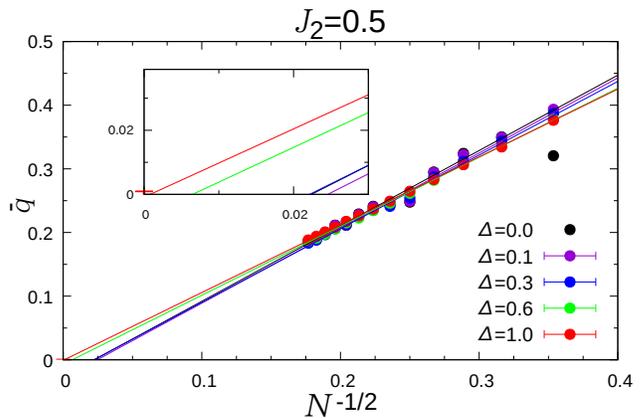}
    \caption{(Color online) The spin freezing parameter $\bar{q}$ plotted versus $1/\sqrt{N}$ for various values of $\Delta$ for $J_2=0.5$. Lines are linear fits of the data. Insets are magnified views of the large-$N$ region. In the fit for $\Delta=0$, the data point of $N=8$ is excluded. 
}
    \label{fig:qb2}
  \end{center}
\end{figure}

\begin{figure}
  \begin{center}
    \includegraphics[clip,width=0.7\hsize]{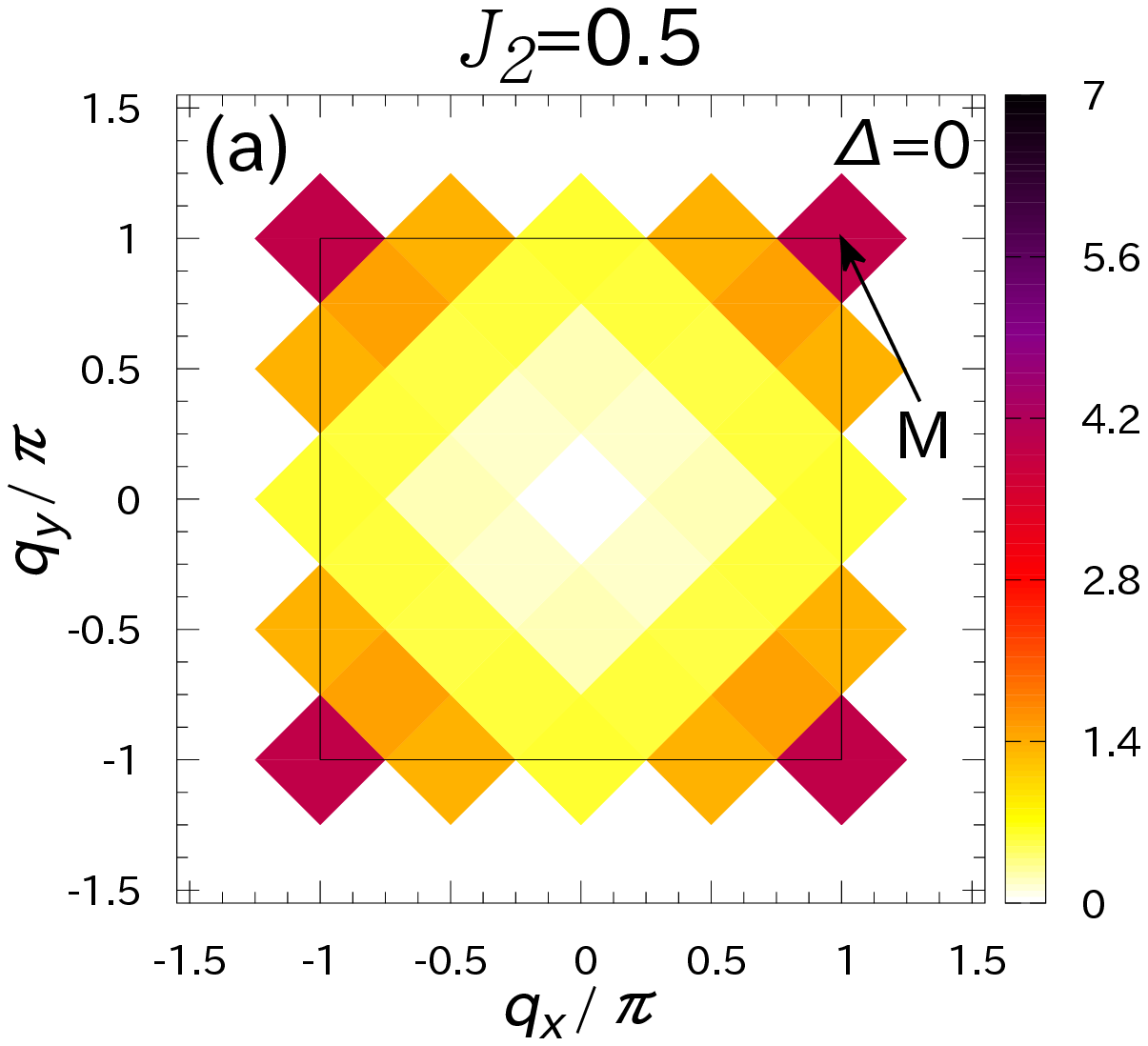}
    \includegraphics[clip,width=0.7\hsize]{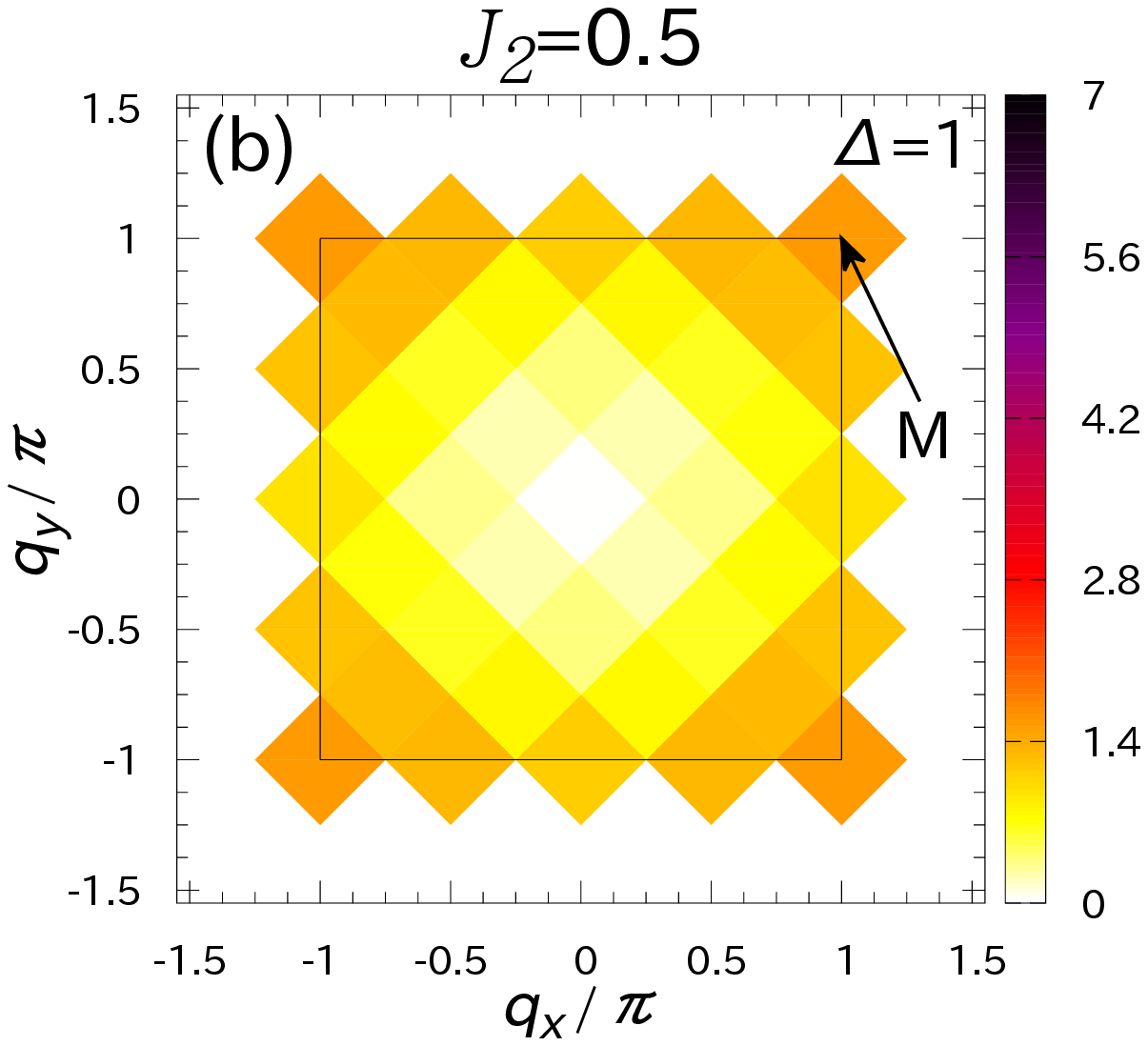}
    \includegraphics[width=\hsize]{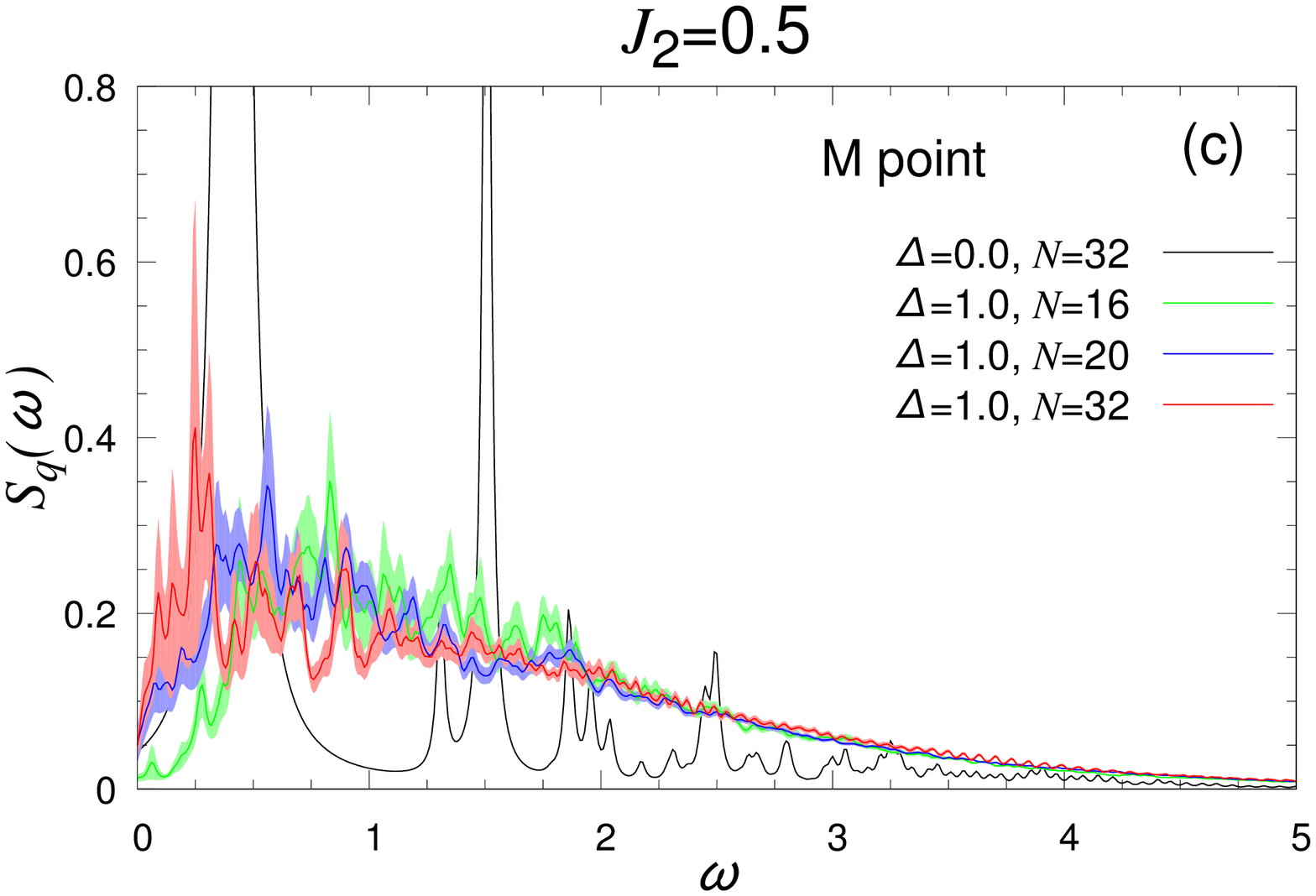}    
\caption{(Color online) Intensity plots of the static spin structure factor $S_{\bm q}$ for $J_2=0.5$: (a) the regular model of $\Delta=0$ and (b) the maximally random model of $\Delta=1$. The lattice size is $N=32$. The solid line depicts the boundary of the first Brillouin zone. (c) The $\omega$-dependence of the dynamical spin structure factor $S_{\bm q}(\omega)$ for the maximally random model of $\Delta=1$ computed at the M point for $J_2=0.5$, as compared with $S_{\bm q}(\omega)$ of the regular model of $\Delta=0$. The lattice size is $N=16$, 20, and 32. Error bars are represented by the width of the data curves.
}
   \label{fig:SSF2}
  \end{center}
\end{figure}
\begin{figure}
  \begin{center}
    \includegraphics[clip,width=\hsize]{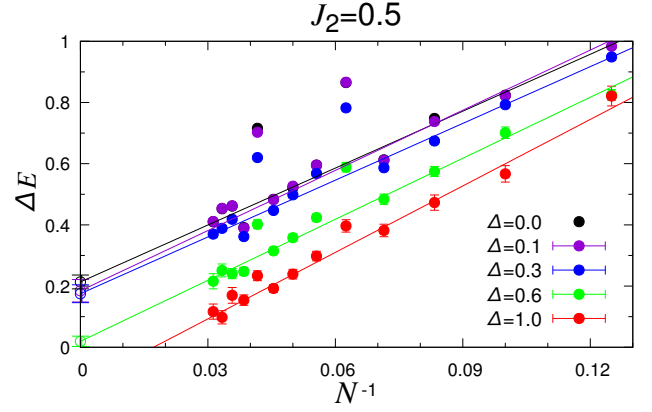}
    \caption{(Color online) The mean spin-gap energy $\Delta E$ plotted versus $1/N$  for various values of $\Delta$ for $J_2=0.5$. The lines are linear fits of the data. The data points of $N=16$ and 24 are excluded in the fit since they largely deviate from other data.
}
   \label{fig:gap2}
  \end{center}
\end{figure}
 Next, we move to the intermediate $J_2$ region of $0.4\leq J_2\leq 0.6$, corresponding to the nonmagnetic ``gapped'' phase of the regular model. To examine the possible magnetic LRO including the spin-glass order,
 we show the size dependence of the freezing parameter $\bar q$ in Fig. \ref{fig:qb2} for $J_2=0.5$. As can be seen from the figure, $\bar q$ is extrapolated to zero within the error bar for any value of $\Delta$, indicating that the ground state in this region is always nonmagnetic. 

 In Figs. \ref{fig:SSF2} (a) and (b), we show  for $J_2=0.5$ the static spin structure factor $S_{\bm q}$ both for the regular case of $\Delta=0$ and for the maximally random case of $\Delta=1$, respectively. In the regular case of $\Delta=0$, a broad peak appears at the M point suggestive of the AF SRO. In the maximally random case of $\Delta=1$, the peak structure is hardly discernible, which is fully consistent with the expected random-singlet state.

 In Fig. \ref{fig:SSF2} (c), we show $S_{\bm q}(\omega)$ for the maximally random case $\Delta=1$, together with that for the regular case $\Delta=0$.  As can be seen from the figure, the data for the random case are much less peaky than the ones of the regular system, with very broad components extending to higher $\omega$. Such a feature is a characteristic of the random-singlet state. \cite{Kawamura,Shimokawa,Uematsu} This resemblance also justifies our identification of the randomness-induced gapless nonmagnetic state observed in the $0.4\leq J_2 \leq 0.6$ region for larger $\Delta$ as the random-singlet state. The marked difference observed in $S_{\bm q}(\omega)$ between $\Delta=0$ and $\Delta=1$ also suggests that the random-singlet state might differ in nature from the nonmagnetic ``gapped'' state realized in the regular system.

 In Fig. \ref{fig:gap2}, we show the size dependence of the spin-gap energy $\Delta E$ for $J_2=0.5$. Interestingly, the spin gap $\Delta E$ is extrapolated to zero in the $N=\infty$ limit, i.e., gapless for stronger randomness of $\Delta>\Delta_c$, while it becomes nonzero, i.e., gapped for weaker randomness of $\Delta<\Delta_c$. The borderline value of $\Delta_c$ is estimated to be $\Delta_c\simeq 0.6$. The gapped nonmagnetic state stabilized for smaller $\Delta$ corresponds to that of the regular model discussed in the literature, while the gapless nonmagnetic state is likely to be the random-singlet state as discussed above. Thus, on increasing the randomness $\Delta$ at $J_2=0.5$, there is a phase transition from the randomness-irrelevant gapped spin-liquid-like state to the randomness-relevant gapless spin-liquid-like state, i.e., the random-singlet state.

 \subsection{Region $0.6\leq J_2 \leq 1$}
\begin{figure}
  \begin{center}
    \includegraphics[clip,width=0.7\hsize]{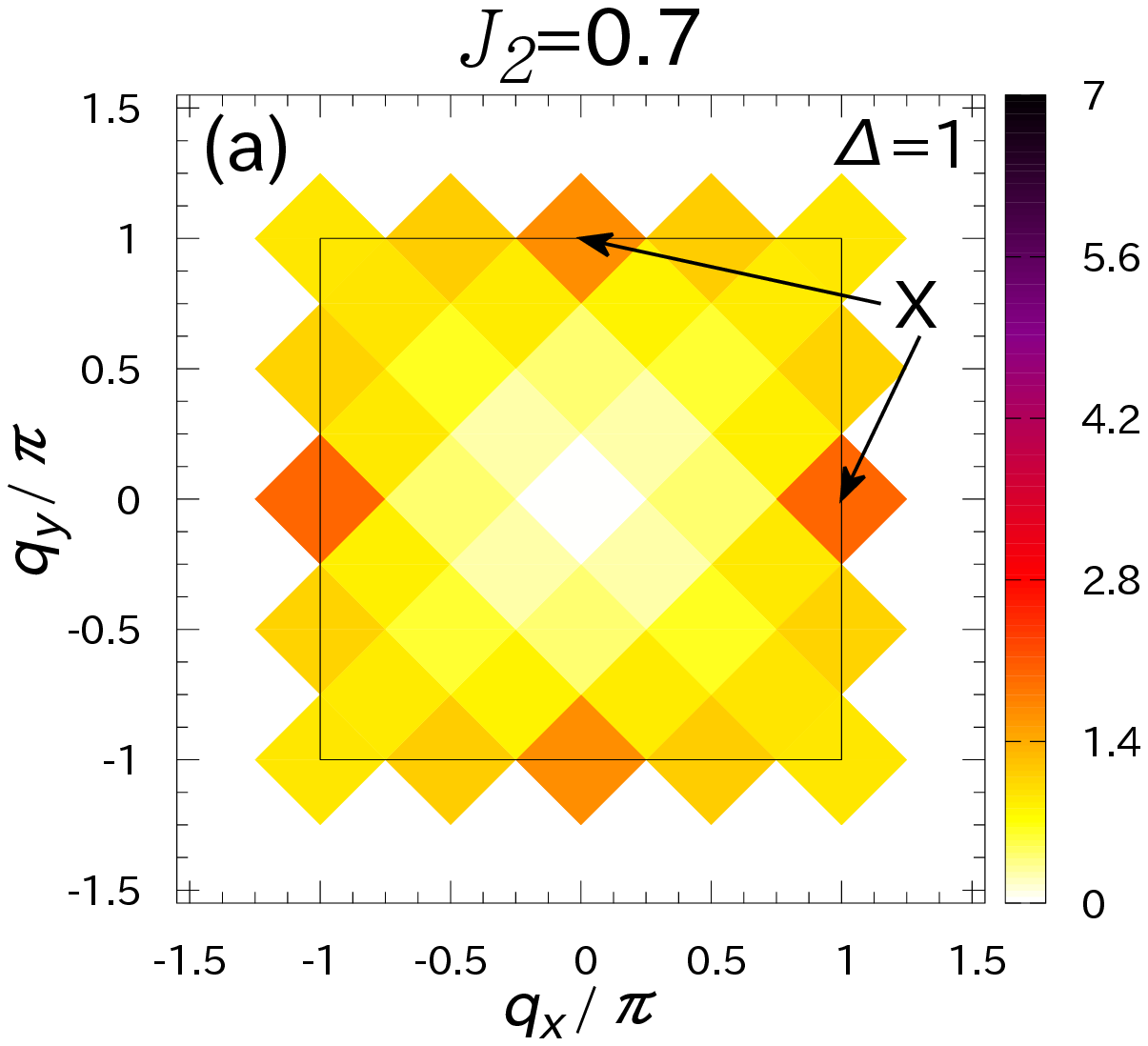}
    \includegraphics[clip,width=\hsize]{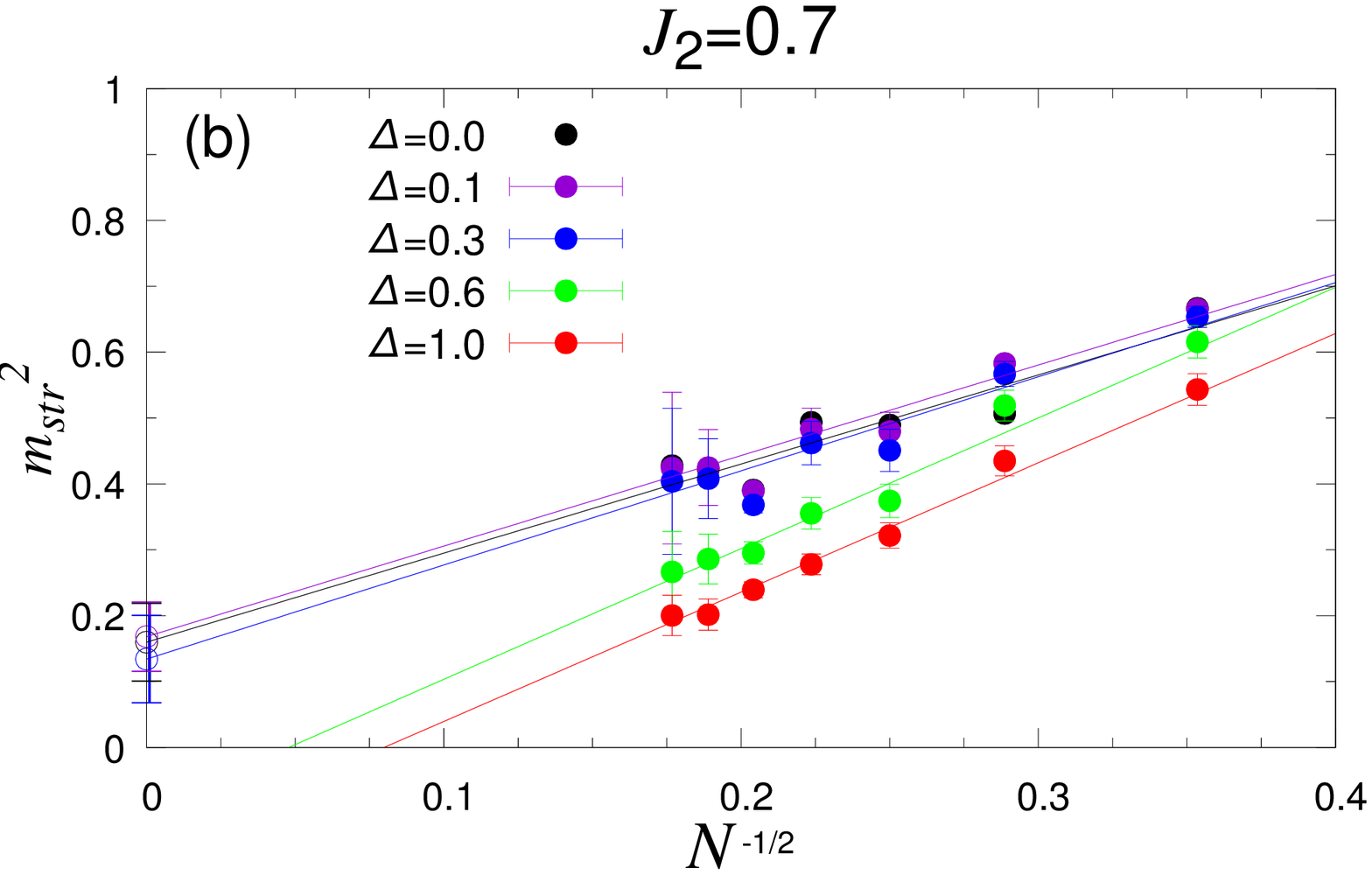}
    \includegraphics[clip,width=\hsize]{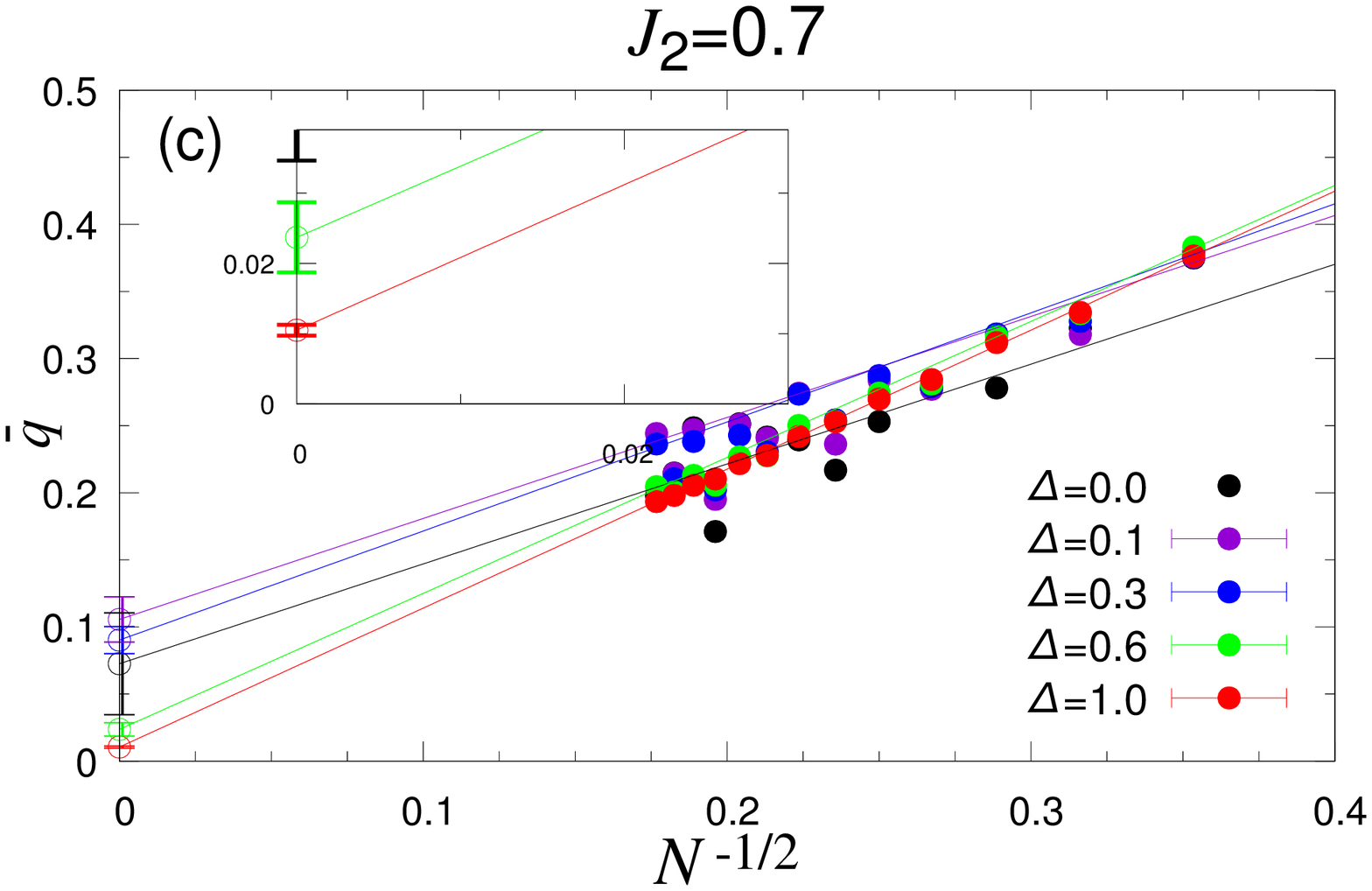}
    \caption{(Color online) (a) Intensity plots of the static spin structure factor $S_{\bm q}$ of the maximally random model of $\Delta=1$ for $J_2=0.7$. The lattice size is $N=32$. The solid line shows the boundary of the first Brillouin zone. (b) The squared stripe order parameter $m_{str}^2$ plotted versus $1/\sqrt{N}$ for various values of the randomness $\Delta$ for $J_2=0.7$. The lines are linear fits of the data. (c) The spin freezing parameter $\bar{q}$ plotted versus  $1/\sqrt{N}$ for various values of $\Delta$ for $J_2=0.7$. The lines are linear fits of the data. The inset is a magnified view of the large-$N$ region.}
    \label{fig:SSF3}
  \end{center}
\end{figure}
\begin{figure}
  \begin{center}
    \includegraphics[width=\hsize]{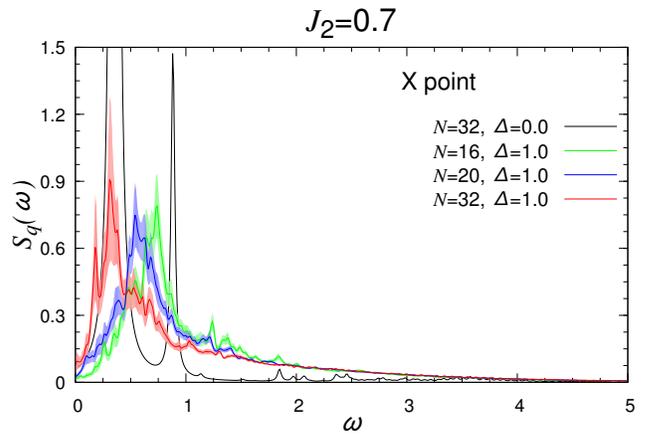}
    \caption{(Color online) The dynamical spin structure factor $S_{\bm q}(\omega)$ computed at the X point for $\Delta=1$ and $J_2=0.7$, as compared with those of the regular model of $\Delta=0$ and $J_2=0.7$. The lattice size is $N=16$, 20, and 32. Error bars are represented by the width of the data curves.}
    \label{fig:DSF3}
  \end{center}
\end{figure}
\begin{figure}
  \begin{center}
    \includegraphics[clip,width=0.7\hsize]{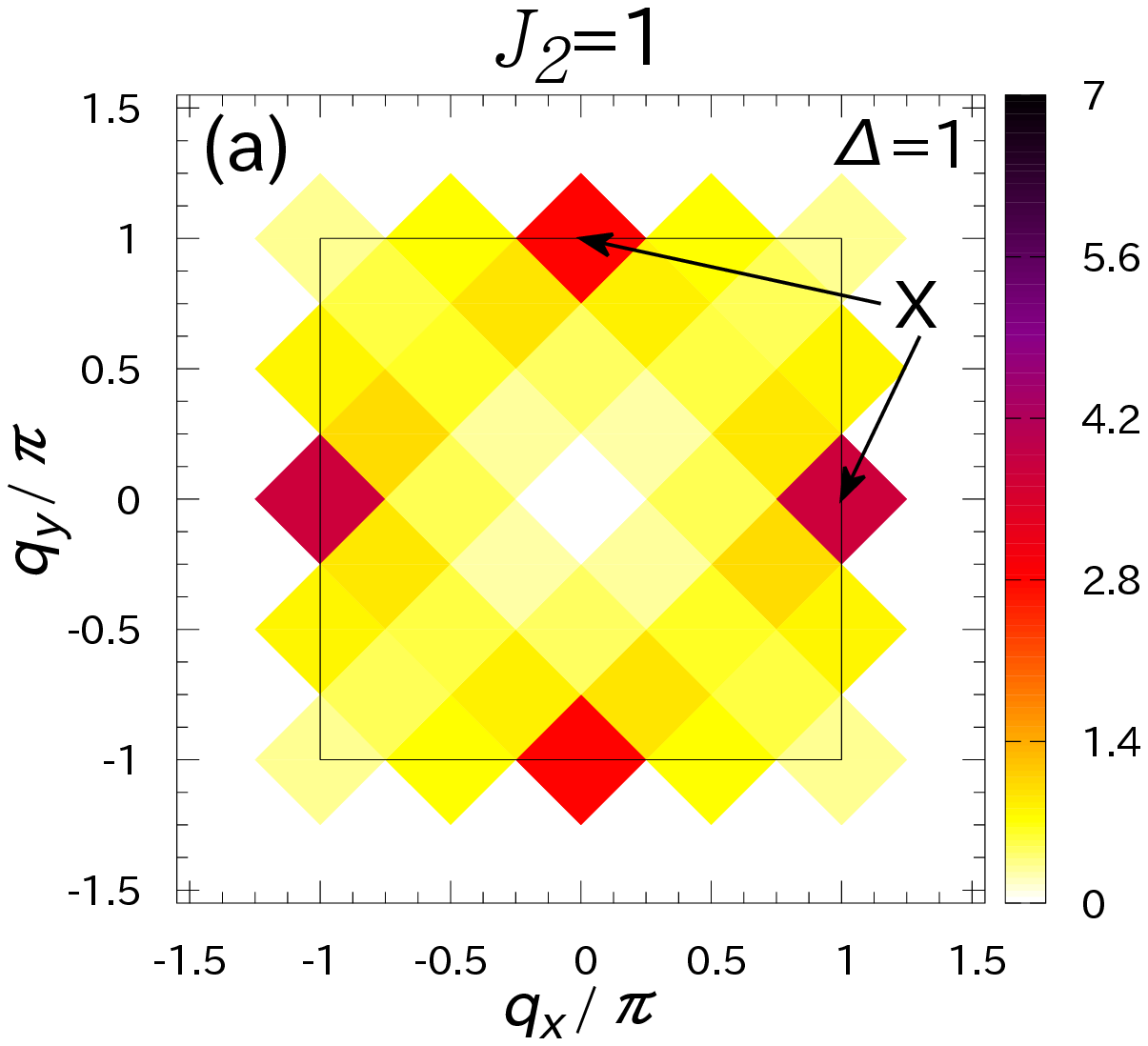}
    \includegraphics[clip,width=\hsize]{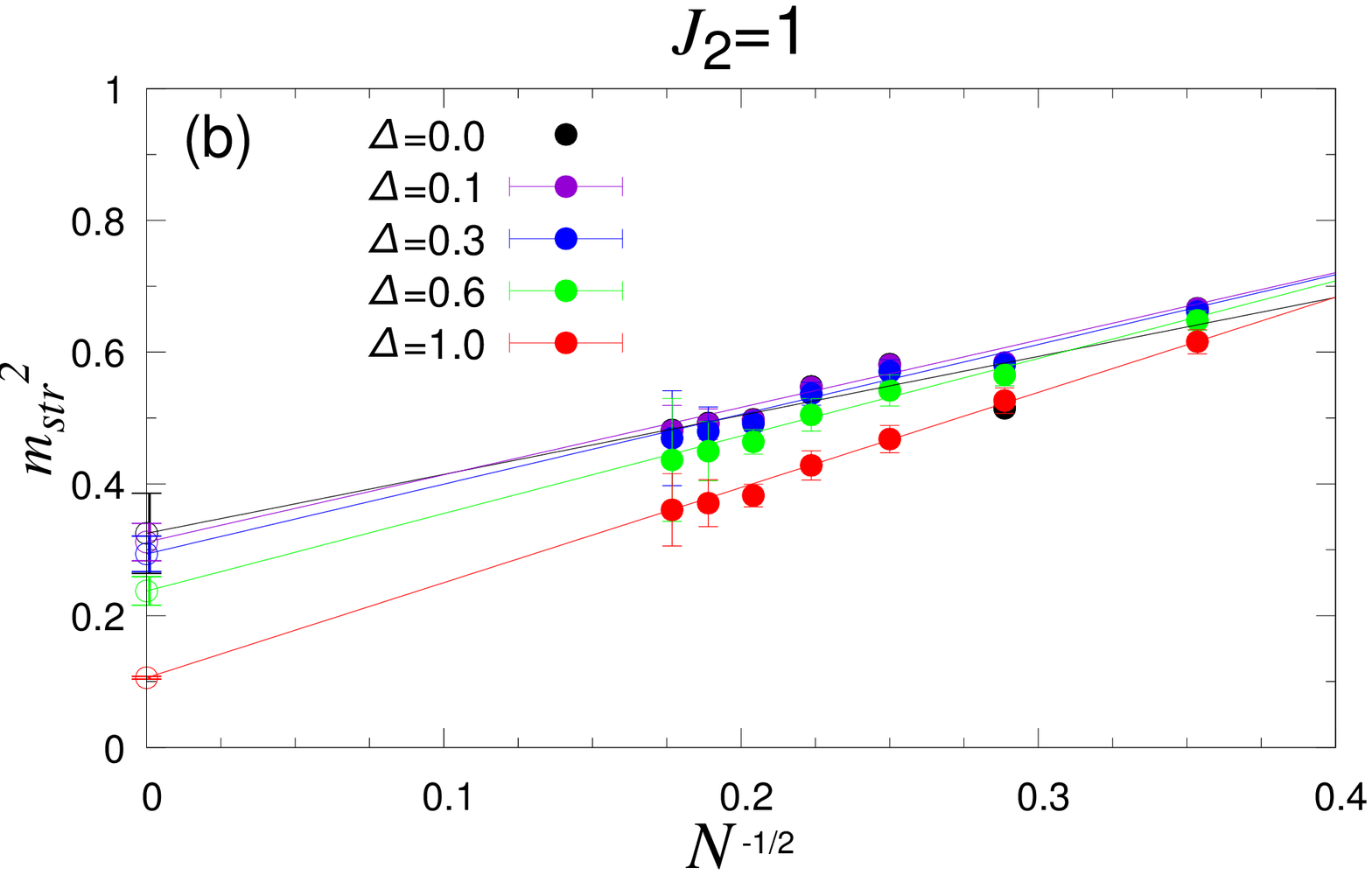}
    \caption{(Color online) (a) Intensity plots of the static spin structure factor $S_{\bm q}$ for $\Delta=1$ and $J_2=1$. The lattice size is $N=32$. The solid line shows the boundary of the first Brillouin zone. (b) The squared stripe order parameter $m_{str}^2$ plotted versus $1/\sqrt{N}$ for various values of the randomness $\Delta$ for $J_2=1$. The lines are linear fits of the data.}
   \label{fig:SSF4}
  \end{center}
\end{figure}

In this subsection, we deal with the larger-$J_2$ region of $0.6\leq J_2\leq 1$, corresponding to the stripe-ordered phase of the regular model. In the small- and intermediate-$J_2$ regions, the randomness-induced state is the nonmagnetic random-singlet state.\cite{Watanabe,Kawamura,Shimokawa,Uematsu} In the region of $0.6\leq J_2\leq 1$, by contrast, we find that the randomness induces the {\it magnetic} state, i.e., the spin-glass state.

 We first consider the case of $J_2=0.7$. In Fig. \ref{fig:SSF3} (a), we show the static spin structure factor $S_{\bm q}$ for the maximally random case of $\Delta=1$. A broad peak corresponding to the stripe-ordered state appears at the X point located at ${\bm q}=(0,\pi)$ and $(\pi,0)$. However, the peak is broad, suggesting that the stripe order here is a SRO.

 This can be checked more quantitatively by computing the magnetic order parameter associated with the stripe order (the X-point order), $m_{str}^2$, defined by
\begin{align}
m_{str}^2 &=\frac{4}{N(N+4)} \left[\sum_{\nu} \sum_{\alpha_\nu}\sum_{i,j\in \alpha_\nu}
\Braket{{\bm S}_i \cdot {\bm S}_j} \right]_J ,
\label{eq:mstr}
\end{align}
where $\nu=1$ and 2 refer to the two distinct types of stripe order.
The computed stripe-order parameter $m_{str}^2$
for various values of $\Delta$ are
given in Fig. \ref{fig:SSF3} (b).

As can be seen from the figure, there exists a finite critical randomness $\Delta_c\simeq 0.5$ beyond which the stripe LRO vanishes. This observation supports our conjecture above that the broad X-point peak observed in $S_q$ shown in Fig. \ref{fig:SSF3} (a) indeed corresponds to the stripe SRO.

 In Fig. \ref{fig:SSF3} (c), we show the size dependence of the freezing parameter $\bar q$ for various randomness $\Delta$. Interestingly, as can be seen from this figure, even at $\Delta>\Delta_c\simeq 0.5$, $\bar q$ is extrapolated to positive nonzero values for any value of $\Delta$, significantly beyond the error bars. For example, at $\Delta=0.6$, an extrapolated value is $\bar q=0.0237\pm 0.0050$, while at $\Delta=1$, it is $\bar q=0.0105 \pm 0.0008$. The result indicates that, at $J_2=0.7$, the stable state for the stronger randomness of $\Delta>\Delta_c\simeq 0.5$ is likely to be a spin glass, rather than the random singlet, although the effective moment associated with the spin-glass order estimated from Fig. \ref{fig:SSF3} is rather small, about one-fifth of the full moment. In fact, this is the distinct occasion that the spin-glass state, instead of the random-singlet state, is stabilized for the stronger randomness in frustrated 2D $s=1/2$ antiferromagnets. \cite{Watanabe,Kawamura,Uematsu} The stabilizing mechanism of the spin-glass state will be discussed in Sec. \ref{sec:summary}.

 In Fig. \ref{fig:DSF3} , we show the $\omega$-dependence of the  dynamical spin structure factor $S_{\bm q}(\omega)$ at the X point computed at $\Delta=1$. We see a rather sharp peak in the small-$\omega$ region coexisting with very broad components with a long tail extending to larger $\omega$, as in the case of $J_2=0.3$ shown in Fig. \ref{fig:SSF1} (b). In fact, this latter feature closely resembles a characteristic of the random-singlet state. \cite{Kawamura,Shimokawa,Uematsu} It would mean that the distinction between the random-singlet state and this spin-glass state is hardly visible from the $\omega$-dependence of $S_{\bm q}(\omega)$.

For $J_2 \geq 1$, the stripe-ordered state remains stable for any $\Delta$. This is demonstrated in Fig. \ref{fig:SSF4}: for the static spin structure factor $S_{\bm{q}}$ at $\Delta = 1$ (a), and for the size dependence of the stripe-order parameter $m_{\rm str}^2$ for various $\Delta$ (b).

\begin{figure}
  \begin{center}
    \includegraphics[clip,width=1\hsize]{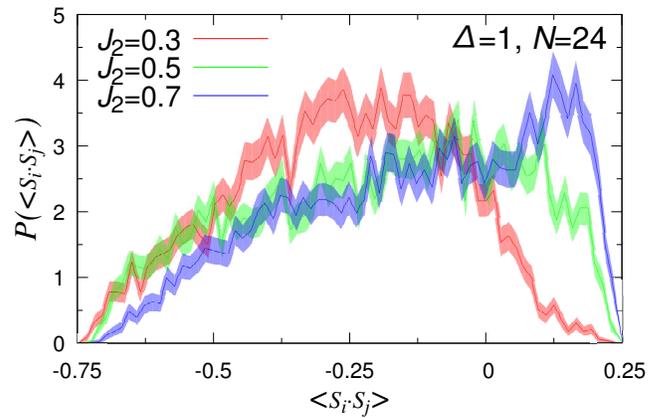}
    \caption{(Color online) The distribution of the nearest-neighbor spin correlation, or the dimer parameter, $\langle {\bm S}_i\cdot {\bm S}_j \rangle$, for $\Delta=1$ and for $J_2=0.3$, 0.5 and 0.7. The lattice size is $N=24$. The system is in the random-singlet state at $J_2=0.3$ and 0.5, but in the spin-glass state for $J_2=0.7$.}
   \label{fig:histo3}
  \end{center}
\end{figure}

The spin correlation in the random-singlet and the spin-glass state have rather similar features in the static and the dynamical spin structure factors, except that the $\bar q$-value is nonzero or not. To get more information about the possible distinction between the two states, we also study the  distribution of the nearest-neighbor spin correlation, or the dimer parameter, defined by $\Braket{ {\bm S}_i\cdot {\bm S}_j }$ over bonds and samples. The computed distribution in the ground state is shown in Fig. \ref{fig:histo3} for the maximally random case of $\Delta=1$ at $J_2=0.3, 0.5$ and 0.7. In the former two cases, the system is in the random-singlet state, while in the latter case, it is in the spin-glass state. For the pure singlet or  the triplet spin pairs, $\Braket{{\bm S}_i\cdot {\bm S}_j}$ takes a value $-3/4$ or $1/4$, respectively, and general $\Braket{ {\bm S}_i\cdot {\bm S}_j }$ values are distributed between these two values.

As can be seen from the figure, $\Braket{{\bm S}_i\cdot {\bm S}_j}$ exhibits a broad distribution spanning between the pure singlet and the pure triplet limits, with a single broad peak located at around $\Braket{{\bm S}_i\cdot {\bm S}_j}\simeq -0.1\sim -0.2$. In the spin-glass state, $\Braket{{\bm S}_i\cdot {\bm S}_j}$ also exhibits a broad distribution, but with a rather clear peak near the triplet end, in contrast to the random-singlet case. The existence of a clear triplet-like peak indicates that a certain fraction of bonds has a ferromagnetic spin correlation in spite of the purely AF interaction in the present model, marking the spin-glass state in distinction with the random-singlet state.

\section{The finite-temperature properties}
\label{sec:finitemp}
\begin{figure}
  \begin{center}
    \begin{tabular}{c}
      \begin{minipage}{0.5\hsize}
        \begin{center}
          \includegraphics[clip,width=\hsize]{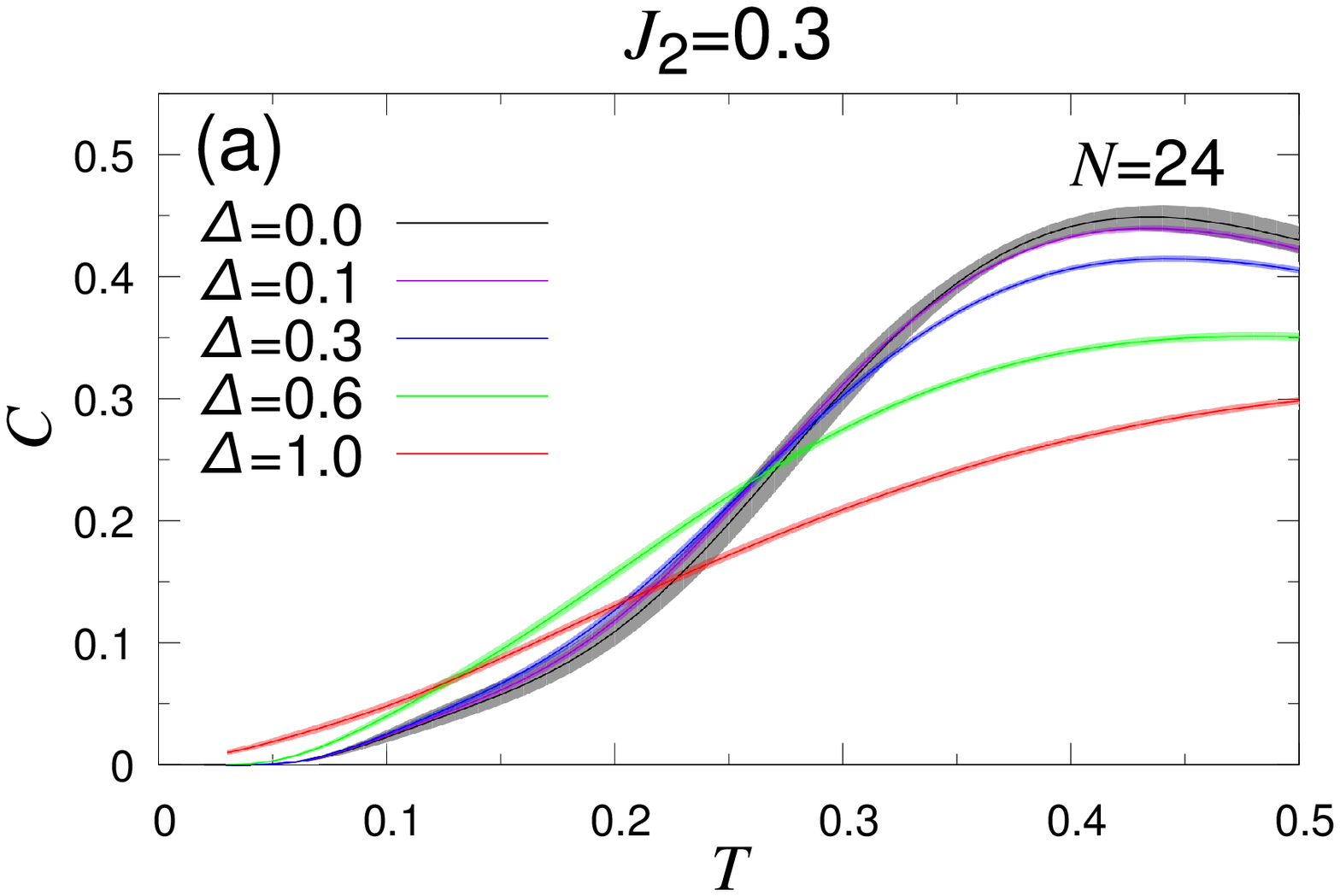}
        \end{center}
      \end{minipage}
      \begin{minipage}{0.5\hsize}
        \begin{center}
          \includegraphics[clip,width=\hsize]{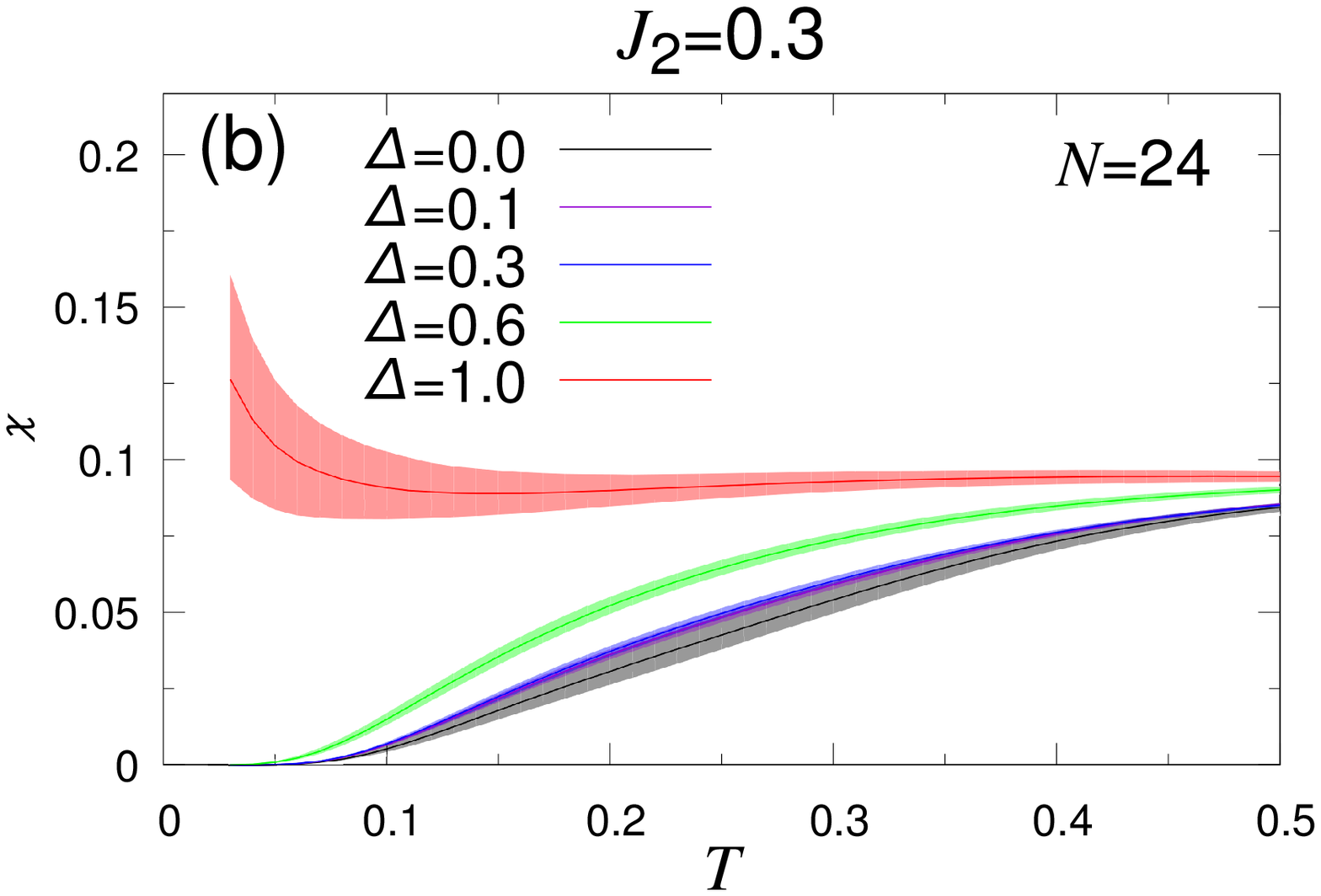}
        \end{center}
      \end{minipage}\\
      \begin{minipage}{0.5\hsize}
        \begin{center}
          \includegraphics[clip,width=\hsize]{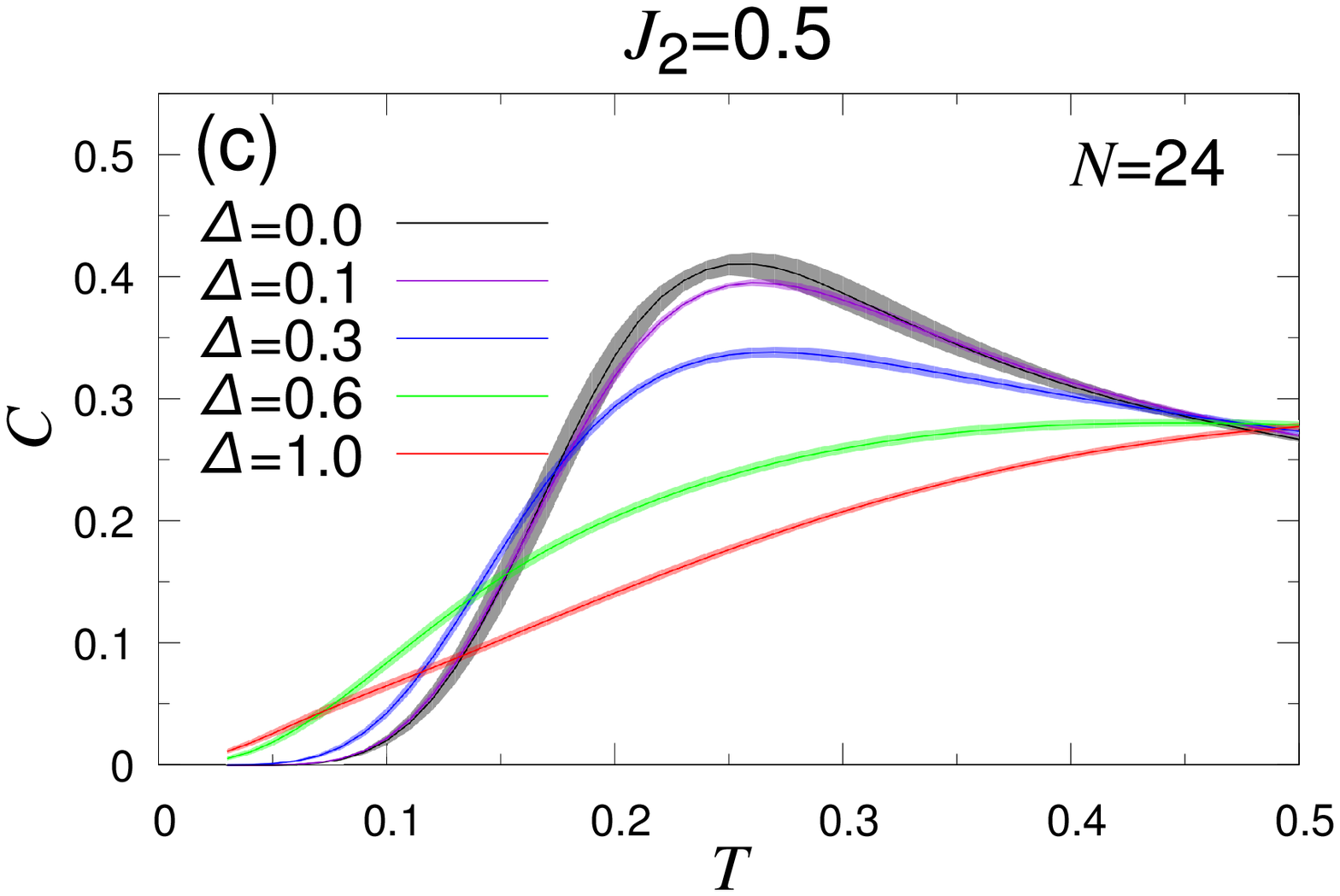}
        \end{center}
      \end{minipage}
      \begin{minipage}{0.5\hsize}
        \begin{center}
          \includegraphics[clip,width=\hsize]{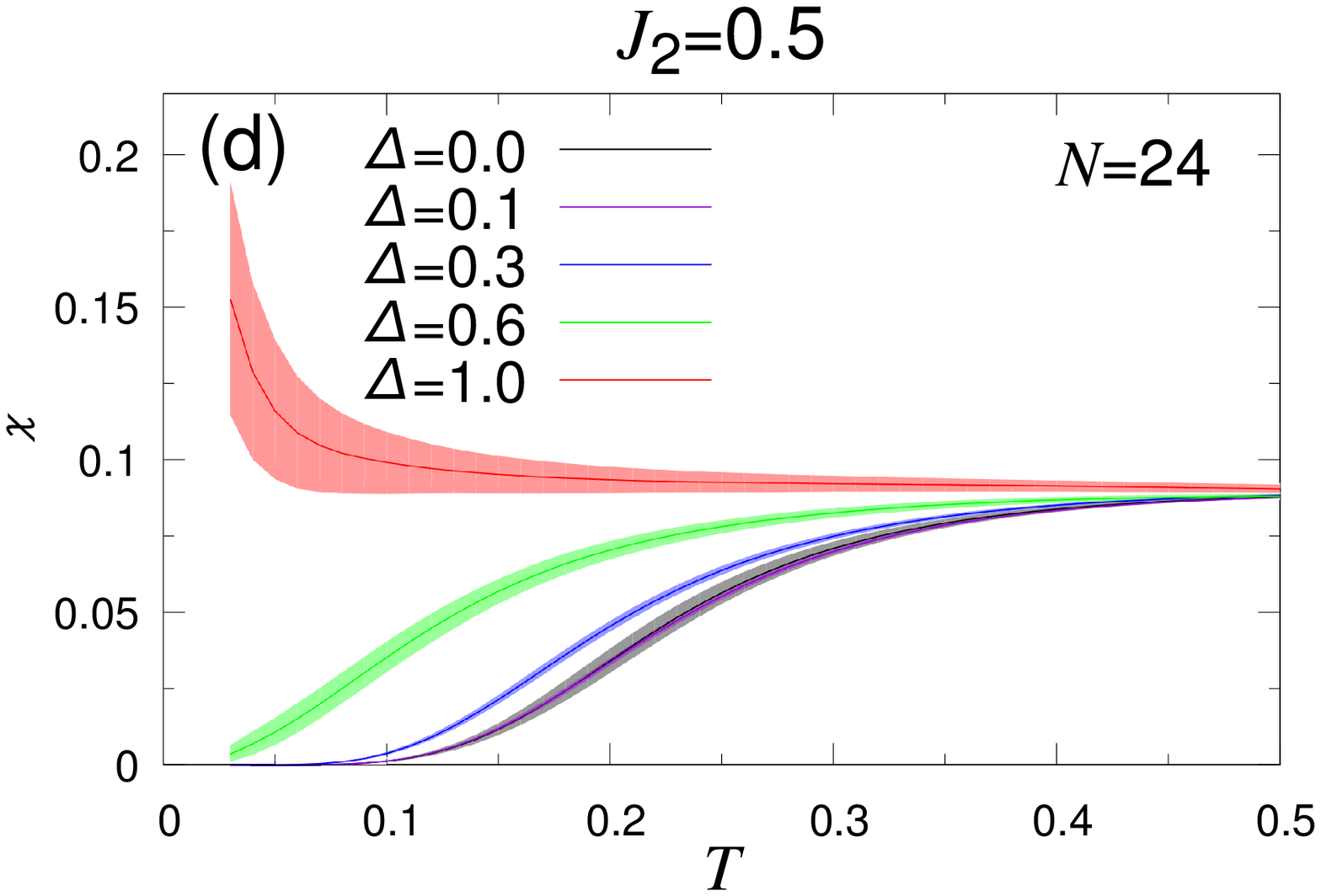}
          \end{center}
      \end{minipage}\\
      \begin{minipage}{0.5\hsize}
        \begin{center}
          \includegraphics[clip,width=\hsize]{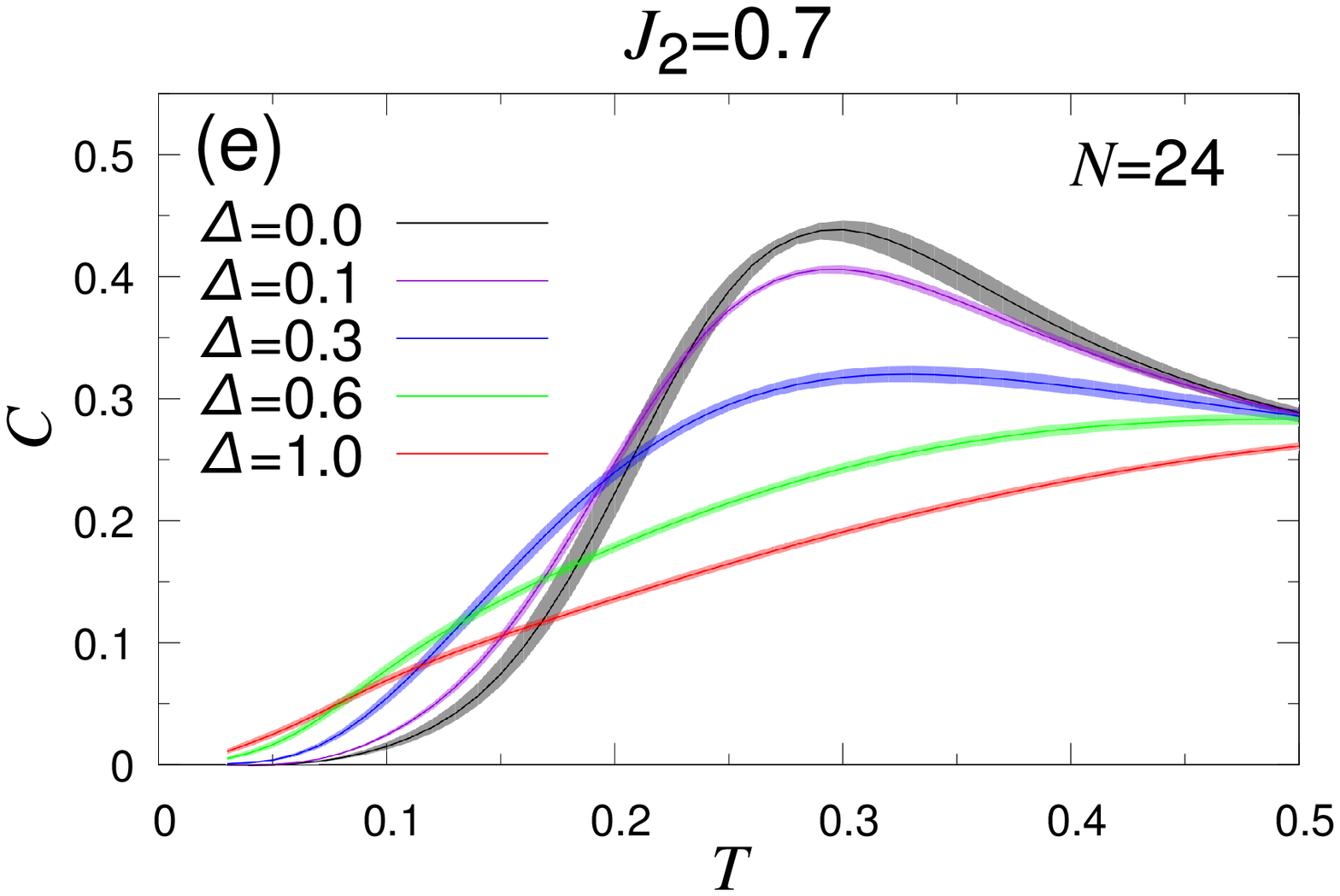}
        \end{center}
      \end{minipage}
      \begin{minipage}{0.5\hsize}
        \begin{center}
          \includegraphics[clip,width=\hsize]{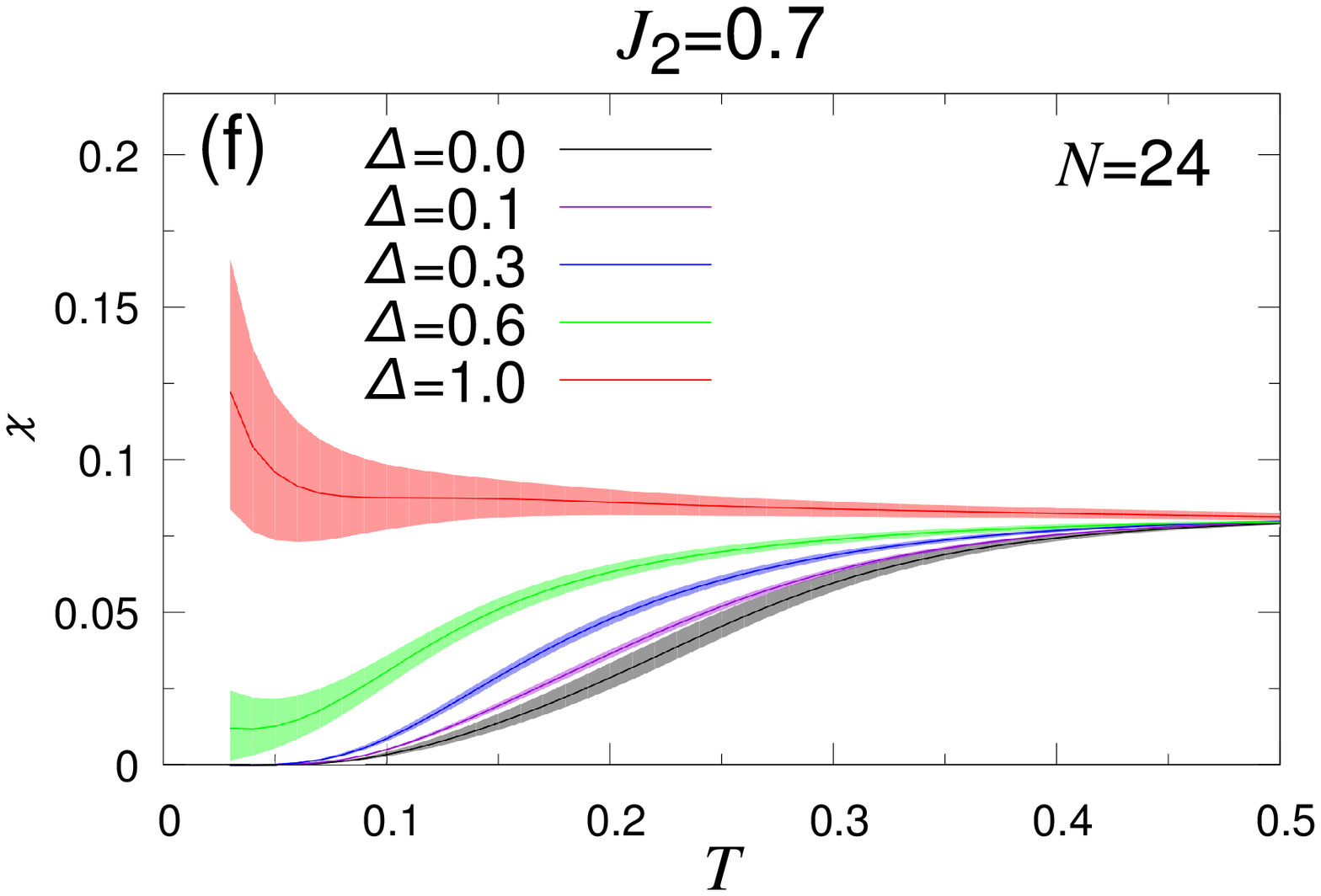}
        \end{center}
      \end{minipage}\\
      \begin{minipage}{0.5\hsize}
        \begin{center}
          \includegraphics[clip,width=\hsize]{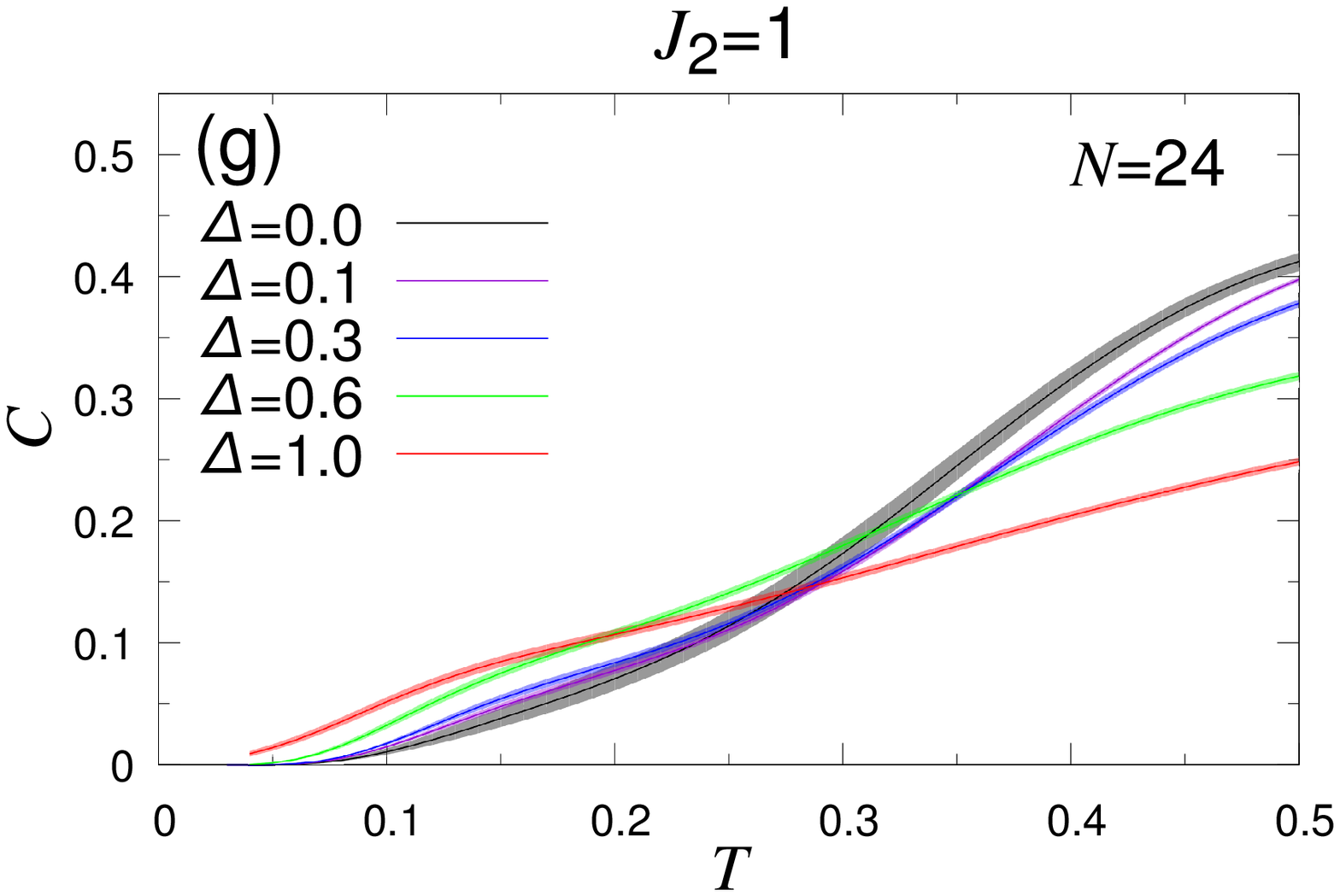}
        \end{center}
      \end{minipage}
      \begin{minipage}{0.5\hsize}
        \begin{center}
          \includegraphics[clip,width=\hsize]{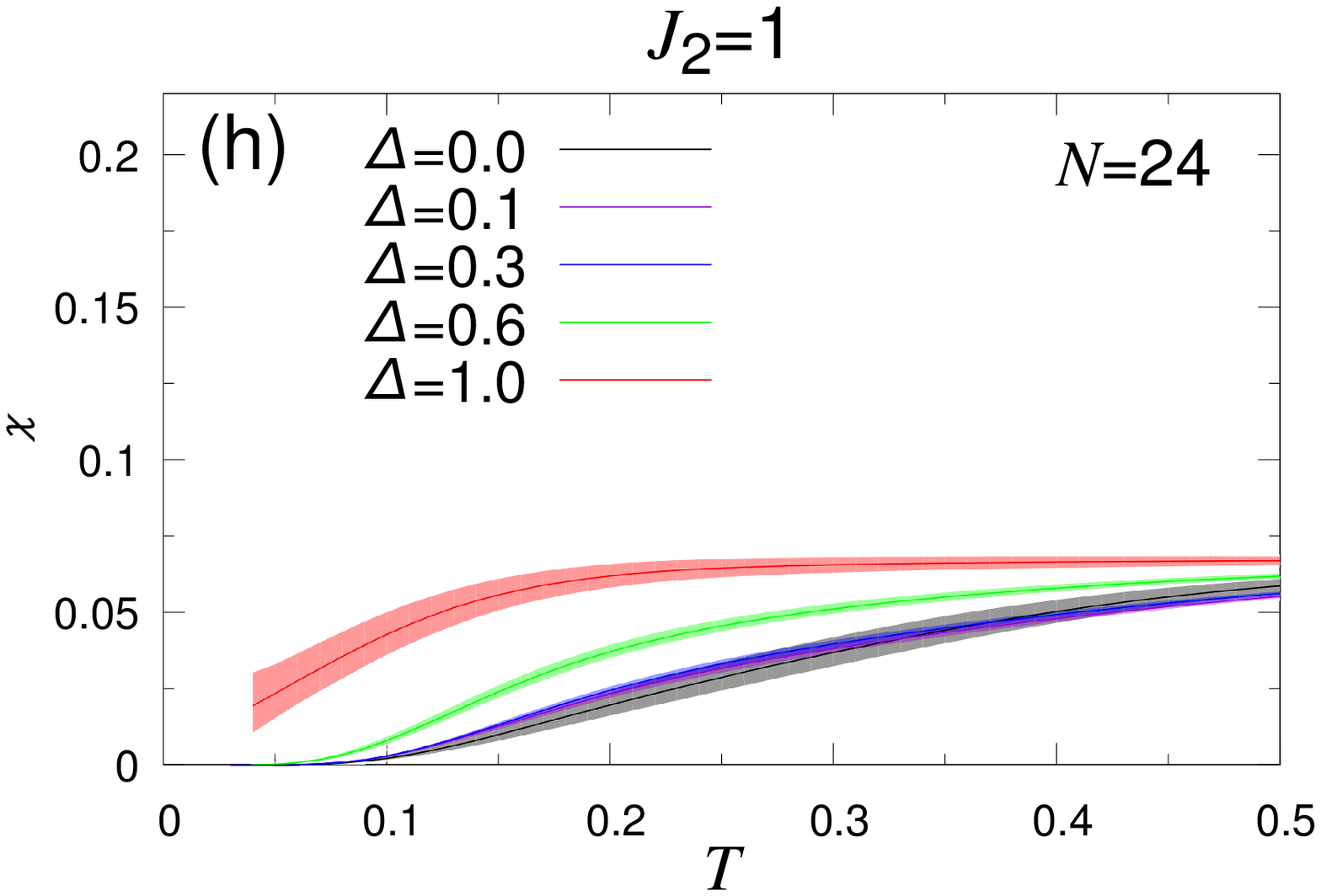}
        \end{center}
      \end{minipage}
    \end{tabular}
    \caption{(Color online) The temperature dependence of the specific heat per spin $C$ [left column], and of the uniform susceptibility per spin $\chi$  [right column], for various values of the randomness $\Delta$ for $J_2=0.3$ (a), (b), $J_2=0.5$ (c), (d),  $J_2=0.7$ (e), (f), and $J_2=1$ (g), (h).}
   \label{fig:finiT}
  \end{center}
\end{figure}

In this section, we investigate the finite-temperature properties of the bond-random $J_1$-$J_2$ Heisenberg model on the square lattice, focusing on the temperature dependence of the specific heat and the uniform susceptibility. To compute the thermal average of these quantities, we employ the Hams--de Raedt method. \cite{HamsRaedt} The results are shown in Figs. \ref{fig:finiT}(a)--\ref{fig:finiT}(h), where the specific heat is given on the left column and the uniform susceptibility on the right column. The temperature dependence of these quantities are given for $L=24$ for various values of randomness $\Delta$, each row corresponding to the case of $J_2=0.3, 0.5, 0.7$ and $1.0$, respectively.

 The ground-state results in the previous section indicate that the system is in the random-singlet state for ($J_2=0.3$, $\Delta=1.0$) and  ($J_2=0.5$, $\Delta=1.0$). 
The low-temperature specific heat there exhibits a $T$-linear behavior, $C\simeq \gamma T$, with $\gamma\simeq 0.4$ at ($J_2=0.3$, $\Delta=1.0$) and $\gamma\simeq 0.6$ at ($J_2=0.5$, $\Delta=1.0$). Such $T$-linear behaviors are fully consistent with the ones observed earlier in the random-singlet state identified for other lattices.
Note that the finite-size effect in the observed $T$-linear behavior is present, but very weak. This is because the random average over various $J_{ij}$ realizations gives rise to low-energy excitations of varying energy scales, including the almost gapless one.

 The $T$-linear nature of the low-temperature specific heat, observed quite robustly in the random-single states identified in a wide variety of randomly frustrated 2D models including the triangular-, \cite{Watanabe} the kagome-, \cite{Kawamura} the $J_1$-$J_2$ honeycomb- \cite{Uematsu} and the $J_1$-$J_2$ square-lattice models suggests an underlying common physical origin. Some time ago, Anderson, Halperin, and Varma presented a phenomenological argument concerning the origin of the $T$-linear low-temperature specific heat widely observed in molecular glasses and spin glasses. \cite{AHV} These authors postulated that, in spin glasses (molecular glasses), the low-energy excitations are borne by the collective flipping of spin clusters (the collective movement of molecule clusters). Due to the random character of the environment surrounding these clusters, their excitation energy $\epsilon$ would exhibit a continuous distribution $\rho(\epsilon)$ with a nonzero weight even in the $\epsilon\rightarrow 0$ limit, i.e., $\rho(\epsilon)>0$ as $\epsilon \rightarrow 0$. This assumption immediately yields the $T$-linear low-$T$ specific heat. If one replaces ``the flipping of local spin clusters'' in spin glasses by, say, ``the singlet-to-triplet excitation of local spin singlets'' or by ``the recombination of singlet pairs (the local resonance of singlets)'' in random-singlet states, the subsequent phenomenological argument leading to the $T$-linear specific heat would be more or less common. This analogy might provide a plausible explanation of the origin of the $T$-linear specific heat robustly observed in the random-singlet states of a wide variety of models.

 The susceptibility exhibits gapless behaviors with a Curie-like tail at lower temperatures both at ($J_2=0.3$, $\Delta=1.0$) and  ($J_2=0.5$, $\Delta=1.0$). Again, such behaviors are fully consistent with the ones observed earlier in the random-single state identified for other lattices.

 By contrast, our ground-state results in the previous section indicate that the system is in the spin-glass state for ($J_2=0.7$, $\Delta=1.0$). The low-temperature specific heat there exhibits a $T$-linear behavior, with the coefficient $\gamma\simeq 0.6$, which is quite similar to the behavior observed in the random-singlet state. Such a similarity indicates that it would be difficult to distinguish the random-singlet state from the spin-glass state solely from the behaviors of the low-$T$ specific heat. In fact, the $T$-linear specific heat has long been known as a representative characteristic of spin glasses.\cite{SGReview}

 The susceptibility at ($J_2=0.7$, $\Delta=1.0$) exhibits a gapless behavior with a Curie-like tail at lower temperatures, again quite similar to the behavior observed in the random-singlet state. A comment should be in order here, however. Although the behavior of the susceptibility in the spin-glass state shown in Fig. \ref{fig:finiT}(f) is certainly quite similar to the one of the random-singlet state shown in Figs. \ref{fig:finiT}(b) and \ref{fig:finiT}(d), care should be taken in comparing these data to real experimental data in the spin-glass state. Namely, while the susceptibility data shown in Fig. \ref{fig:finiT} are fully equilibrated ones, a full equilibration is usually impossible in the spin-glass state  in real experiments, simply because spins exhibit extremely slow glassy dynamics there. This should be contrasted to the case of the random-singlet state where spins are expected to rapidly fluctuate down to zero temperature and a full equilibration is not hard. In case of real spin glasses, the associated slow spin dynamics would easily drive the system out of equilibrium at low temperatures, leading to the dynamical spin freezing often accompanied by the susceptibility cusp. 
Thus, in real experimental situations, the random-singlet state and the spin-glass state would easily be distinguishable via the behavior of the magnetic susceptibility, even if the fully equilibrated data look similar.

 In case of weaker randomness, the ground-state analysis in the previous section has revealed the three phases. Two are magnetic, i.e., the two-sublattice AF state and the stripe-ordered state, and one is nonmagnetic, i.e., a gapped spin-liquid-like state. In case of the two-sublattice AF state corresponding to ($J_2=0.3$, $\Delta\leq 0.6$), the specific heat exhibits a stronger curvature at low temperatures, apparently consistent with the $T^2$ behavior expected from the spin-wave analysis, while the susceptibility vanishes in the $T\rightarrow 0$ limit. In case of the stripe-ordered state corresponding to ($J_2=0.7$, $\Delta\leq 0.3$) and ($J_2=1.0$, $\Delta\leq 0.6$), more or less similar behaviors are found both in the specific heat and the susceptibility.

 In case of the gapped nonmagnetic state corresponding to ($J_2=0.5$, $\Delta\leq 0.3$), the specific heat exhibits stronger curvature at low temperatures, while the susceptibility vanishes in the $T\rightarrow 0$ limit, consistently with a nonzero spin gap. As mentioned in Sec. \ref{sec:intro}, promising candidates of the gapped phase stabilized in this regime for weaker randomness might be the plaquette or the columnar-dimer states, each of which spontaneously breaks the four-fold degeneracy of the ground state.  In 2D, such a spontaneous symmetry breaking might accompany a finite-$T$ transition with a divergent specific heat, possibly lying in the universality class of the four-state clock model,\cite{Suzuki} as was pointed out in Refs. \onlinecite{Zhitomirsky,Becca-Mila,Takano}. As can be seen from Fig. \ref{fig:finiT} (a), however, the computed specific heat does not exhibit a sharp peak suggestive of such a phase transition. Of course, this might simply be due to the fact that our lattice size $N=24$ is too small to detect such a peak.

\section{Summary and discussion}
\label{sec:summary}

 Both the ground-state and the finite-temperature properties of the bond-random $s=1/2$ $J_1$-$J_2$ Heisenberg model on the square lattice are investigated by means of the ED and the Hams--de Raedt methods. The ground-state phase diagram is constructed in the randomness ($\Delta$) versus the frustration ($J_1/J_2$) plane to obtain insight into the role of randomness and frustration in stabilizing various phases. In the phase diagram, we found two types of randomness-induced states, i.e., the random-singlet state and the spin-glass state. The width of the parameter region where the random-singlet appears is slightly narrow compared to the one of the corresponding honeycomb-lattice model.

\begin{figure}
  \begin{center}
    \includegraphics[width=0.7\hsize]{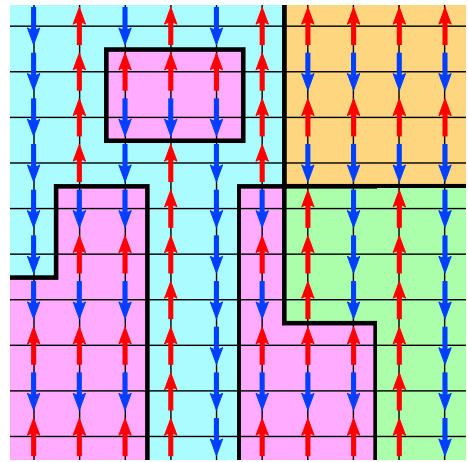}
    \caption{(Color online) Illustration of the spin-glass state of the present model stabilized in the parameter region of the stronger randomness (larger $\Delta$) and the intermediate $J_2$ where the stripe-ordered state is stabilized in the regular limit. This spin-glass state is a random domain state consisting of two distinct types of stripe-ordered states possible in the regular limit, the vertical and the horizontal stripes, which are associated with a twofold directional symmetry of the square lattice.}
   \label{fig:SGpict}
  \end{center}
\end{figure}

 Our present results compare favorably with the recent experimental results on the square-lattice mixed-crystal antiferromagnet Sr$_2$Cu(Te$_{1-x}$W$_{x}$)O$_6$.\cite{Mustonen,Watanabe-Tanaka,Mustonen2} For $x=0.5$, this compound is found to exhibit gapless QSL-like behaviors without any static spin order down to 19 mK, accompanied by the $T$-linear low-$T$ specific heat and the gapless susceptibility with a Curie tail. When we use the experimentally estimated Curie-Weiss temperature $\Theta_{CW}\simeq71$ K as the characteristic temperature scale, we get the $T$-linear term of $\gamma\sim70$ mJ/molK$^2$, which is not far from the experimentally observed $T$-linear term of $\gamma\sim50$ mJ/molK$^2$. As mentioned in Refs. \onlinecite{Mustonen,Watanabe-Tanaka,Mustonen2}, the exchange interaction in this material is of the $J_1$-$J_2$-type with a considerable amount of structural disorder caused by the random distribution of tellurium and tungsten. For the two end compounds, Sr$_2$CuTeO$_6$ and Sr$_2$CuWO$_6$, the ratio $J_2/J_1$ is estimated to be $J_2/J_1\sim 0$ for Sr$_2$CuTeO$_6$, and $J_2/J_1\sim4$-$8$ for Sr$_2$CuWO$_6$. \cite{Mustonen,Watanabe-Tanaka,Mustonen2,Babkevich,Walker} Although the ratio $J_2/J_1$ for Sr$_2$Cu(Te$_{1-x}$W$_{x}$)O$_6$ is not known precisely, the existence of a significant amount of frustration and exchange randomness is likely to locate Sr$_2$Cu(Te$_{0.5}$W$_{0.5}$)O$_6$ lying in the random-singlet state. Hence, a very good possibility exists that the experimentally observed gapless QSL-like behavior of Sr$_2$Cu(Te$_{1-x}$W$_{x}$)O$_6$ might indeed be that of the random-singlet state as was discussed in Refs. \onlinecite{Mustonen,Watanabe-Tanaka,Mustonen2}.

 In the present square-lattice model, unlike the previously studied randomly frustrated 2D models, \cite{Watanabe,Kawamura,Uematsu} the spin-glass state is stabilized in the specific region of the phase diagram where the stripe-ordered state is stabilized in the regular limit. We note that the stripe-ordered state has a twofold degeneracy associated with the $\frac{\pi}{2}$-rotation symmetry of the lattice. We deduce that these two degrees of freedom associated with the stripe order might be essential in inducing the stable spin-glass state. Under the random \{ $J_{ij}$ \} environment, the local energy is different between the horizontal and the vertical stripes depending on the local \{ $J_{ij}$ \} background so that either of the horizontal or the vertical stripes would be favored locally: the situation is illustrated in Fig. \ref{fig:SGpict}. Then, the random domain state consisting of the horizontal and the vertical stripes adjusted to the local \{ $J_{ij}$ \} could lower the energy and be stabilized, relative to the random-singlet state which possesses no such lattice-symmetry-related degrees of freedom. The spin-glass state stabilized in the square-lattice model might essentially be such a random domain state of stripes.

 To test the validity of this conjecture, we examine the possible stabilization of the spin-glass state in another model, i.e., the $s=1/2$ $J_1$-$J_2$-$J_3$ Heisenberg model on the honeycomb lattice (with $J_2=J_3$), where the same mechanism as discussed above is expected. In the regular case of this model, it has been known that the stripe-ordered state with the three-fold degeneracy is stabilized as the ground state for $J_2=J_3\gtrsim0.6$.\cite{Lauchli,Cabra,Reuther,Oitmaa,Li-Bishop} As can be seen from Figs. \ref{fig:J1J2J3}(a) and \ref{fig:J1J2J3}(b) shown in Appendix A, the introduction of the randomness to the stripe-ordered state of the regular model eventually induces the spin-glass state, just as in the case of the square-lattice model. 
Such an observation also supports our conjecture above.

We note that the phase boundary between the stripe-ordered state and the spin-glass state drawn in Fig. \ref{fig:SGpict} might significantly be modified. Indeed, in the thermodynamic limit, even an infinitesimal randomness might destabilize the stripe-ordered state, leading to the spin-glass state due to the ``random-field effect.'' Namely, as discussed in Ref. \onlinecite{Kimchi}, the bond randomness is expected to serve as a random-field conjugate to the underlying stripe order, and the Imry-Ma argument suggesting that the ordered state is unstable against an arbitrarily weak random field conjugate to the order parameter would operate.\cite{Imry-Ma,Binder,Villain} Very small system sizes available in the present calculation tends to mask such a random-field effect operating even for an infinitesimal randomness in the thermodynamic limit. Much larger system sizes are required to directly prove such a theoretical expectation.

 In experiments on Sr$_2$Cu(Te$_{0.5}$W$_{0.5}$)O$_6$, no spin-glass state has been observed so far, though the QSL-like state, most probably the random-singlet state, has been observed. The reason of this might be, at least partially, that the expected spin-glass state is nothing but the random stripe-domain state so that it might look like the ``dirty'' stripe-domain or columnar state, especially when the randomness is weak. In this context, it should be remembered that, although in our present model we have assumed for simplicity that  both $J_1$ and $J_2$ obey a common randomness distribution, the experimental situation in Sr$_2$Cu(Te$_{1-x}$W$_{x}$)O$_6$ might considerably be different. Namely, in Sr$_2$Cu(Te$_{1-x}$W$_{x}$)O$_6$, $J_1$ is borne primarily by Te ($J_2/J_1$ was estimated to be $J_2/J_1\sim 0$ in the end compound Sr$_2$CuTeO$_6$), while $J_2$ primarily by W  ($J_2/J_1$ was estimated to be $J_2/J_1\sim4$-$8$ in the other end compound Sr$_2$CuWO$_6$), so that the $J_2/J_1$ is expected to be highly correlated with the extent of the randomness. In other words, for $J_1$-dominant or $J_2$-dominant systems, the randomness cannot be too strong.


In any case, our present calculation on the random square-lattice model has demonstrated that the randomness-induced gapless QSL-like state, the random-singlet state, prevails in quantum magnets on a variety of frustrated lattices, including the triangular, the kagome, the $J_1$-$J_2$ honeycomb, and the $J_1$-$J_2$ square lattices. We have also clarified that, under certain circumstances, the spin-glass state could be stabilized as a random domain state associated with the lattice-directional degeneracy of the underlying magnetic order.

\begin{acknowledgements}
 The authors wish to thank I. Terasaki, T. Shimokawa, and H. Koushiro for valuable discussions. This study was supported by JSPS KAKENHI Grant No. JP25247064. Our code was based on TITPACK Ver.2 coded by H. Nishimori. We are thankful to ISSP, the University of Tokyo, and to YITP, Kyoto University, for providing us with CPU time.
\end{acknowledgements}

 \appendix
 \section{Random $J_1$-$J_2$-$J_3$ Heisenberg model on the honeycomb lattice}
\begin{figure}
  \begin{center}
    \includegraphics[clip,width=\hsize]{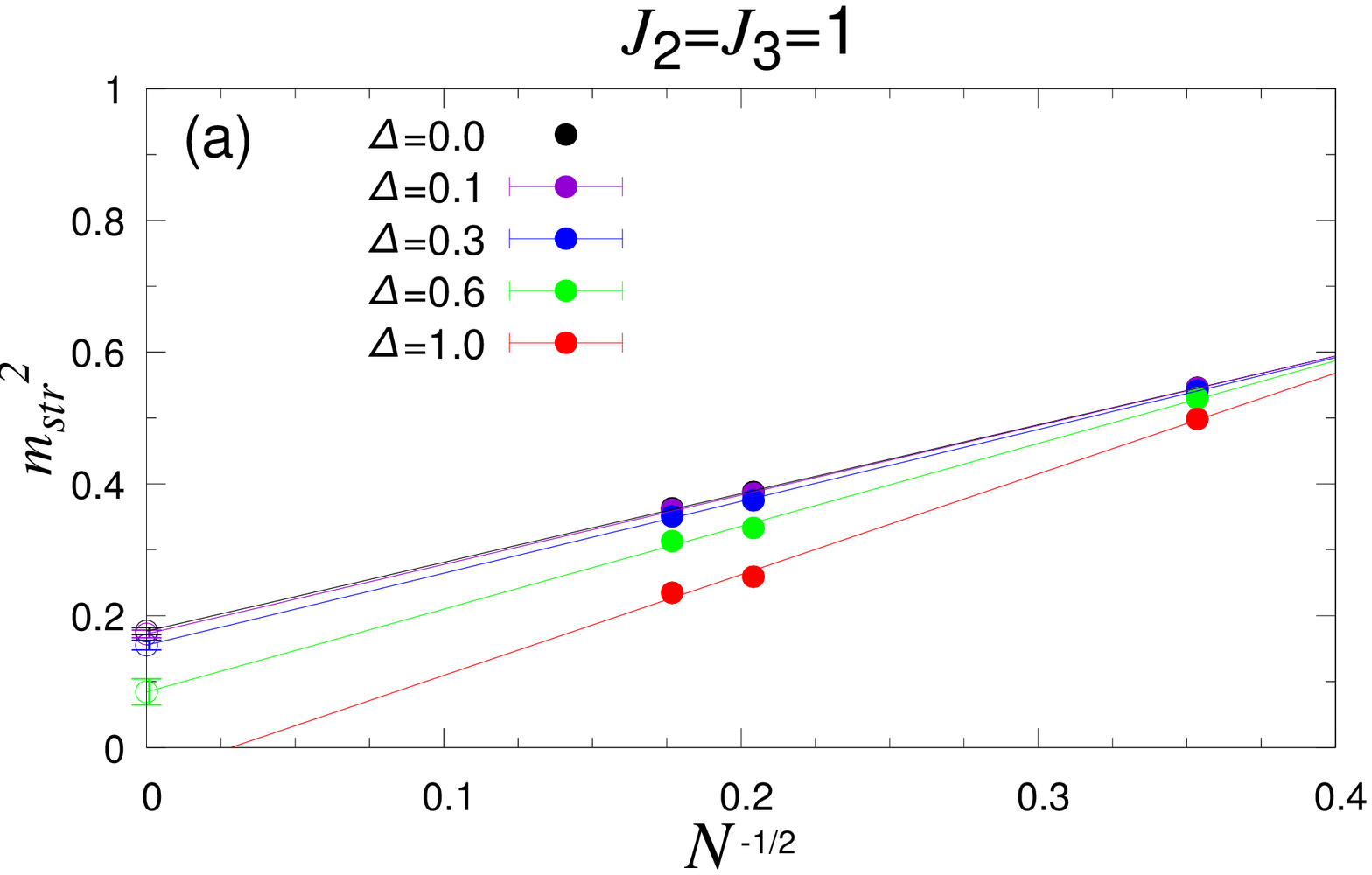}
    \includegraphics[clip,width=\hsize]{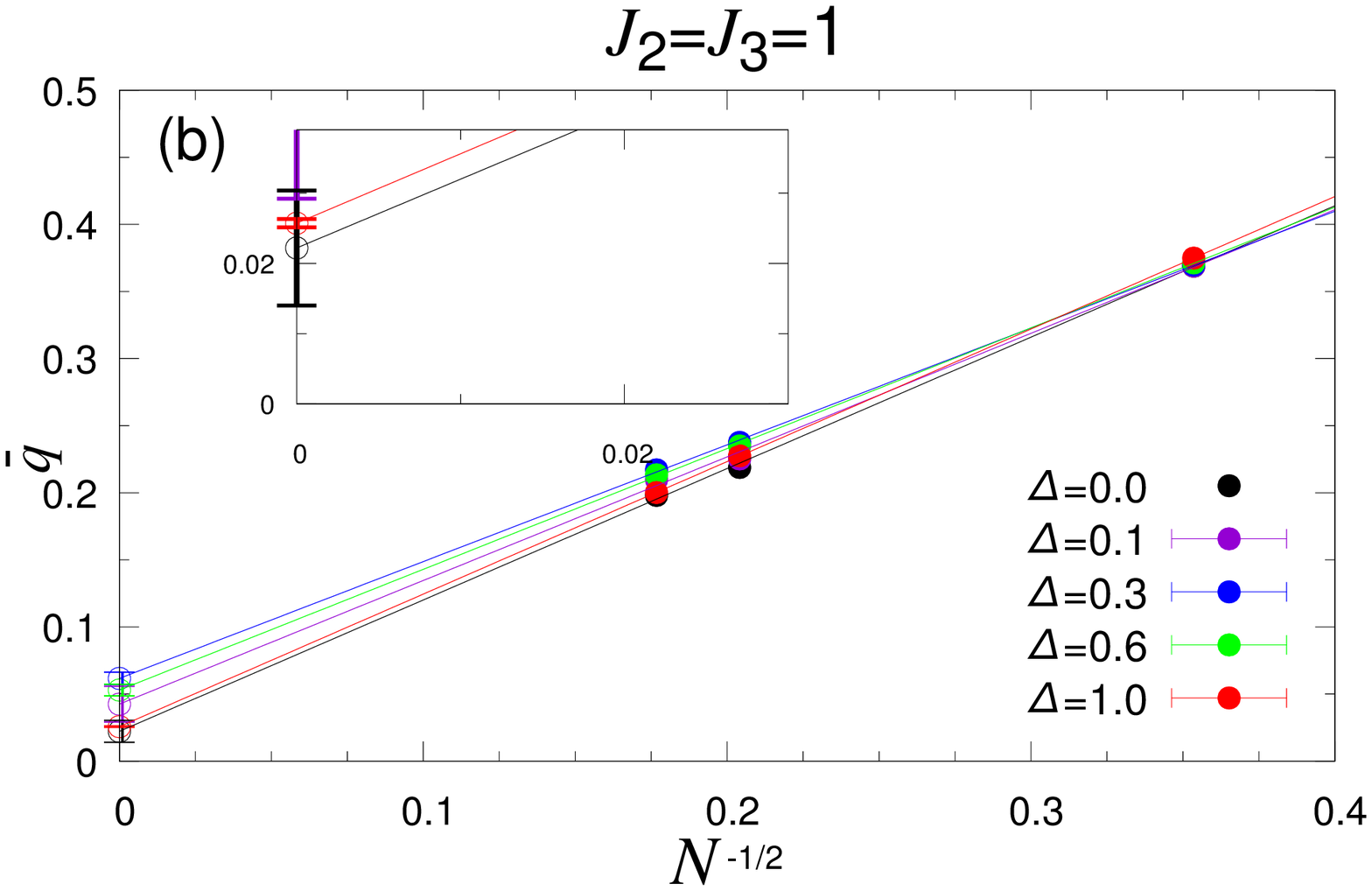}
    \caption{(Color online) (a) The squared stripe-order parameter $m_{str}^2$ and (b) the spin freezing parameter $\bar{q}$ of the bond-random $s=1/2$ $J_1$-$J_2$-$J_3$ Heisenberg model on the honeycomb lattice ($J_2=J_3=1$) plotted versus $1/\sqrt{N}$ for various values of $\Delta$. The lines are linear fits of the data. The inset of (b) is a magnified view of the large-$N$ region.}
    \label{fig:J1J2J3}
  \end{center}
\end{figure}

In this Appendix, we investigate the ground-state properties of the bond-random $s=1/2$ $J_1$-$J_2$-$J_3$ Heisenberg antiferromagnet on the honeycomb lattice, where $J_1$, $J_2$, and $J_3$ are the nearest-, the next-nearest-, and the next-next-nearest-neighbor antiferromagnetic interactions, respectively. For the regular version of the model, it was reported that the stripe-ordered state associated with the threefold degeneracy of the honeycomb lattice is stabilized for $J_2=J_3\gtrsim0.6$. \cite{Lauchli,Cabra,Reuther,Oitmaa,Li-Bishop} Since our interest here is whether the spin-glass state is ever stabilized upon introducing the randomness into the stripe-ordered state, we focus here only on the $\Delta$-dependence of the ground-state properties for fixed $J_2=J_3=1$. The randomness is introduced in the same manner as has been done for the square-lattice model in Sec. II, i.e., a common form of the uniform distribution assumed for all $J_1$, $J_2$, and $J_3$. 

In Fig. \ref{fig:J1J2J3}, we show the $\Delta$-dependence of (a) the squared stripe order parameter $m_{str}^2$ and (b) the spin freezing parameter $\bar{q}$ at $J_2=J_3=1$. As can be seen from Fig. \ref{fig:J1J2J3} (a), the stripe order vanishes in the strongly random region of $\Delta\gtrsim 0.8$, while $\bar{q}$ extrapolated to the thermodynamic limit has a significantly positive value beyond the  error bar for all $\Delta$. These indicate that the randomness destabilizes the stripe-ordered state, and induces the spin-glass state rather than the random-singlet state, just as in the case of the $J_1$-$J_2$ model on the square lattice of $0.6\lesssim J_2\lesssim 1.0$ treated in the main text.

\end{document}